\documentclass{emulateapj} 
\usepackage{graphicx}
\usepackage{amsmath}   
\usepackage{float}
\usepackage{url}
\usepackage{color}
\usepackage{amssymb }
\usepackage{epsfig}
\usepackage{epstopdf}
\usepackage{xfrac}
\usepackage{gensymb}
\usepackage{wasysym}
\usepackage{natbib}
\graphicspath{{./}}

\allowdisplaybreaks[1]

\newcommand{\msun}{\ensuremath{\,M_\odot}}
\newcommand{\rsun}{\ensuremath{\,R_\odot}}

\newcommand{\rp}{\ensuremath{\,R_{\rm P}}}
\newcommand{\aplan}{\ensuremath{\,a_{\rm P}}}
\newcommand{\rs}{\ensuremath{\,R_{\rm S}}}

\newcommand{\pc}{\rm {pc}}

\newcommand{\wasp}{{\rm WASP-103}}
\newcommand{\planet}{{\rm WASP-103{b}}}

\newcommand{\ecosw}{\ensuremath{\sqrt{e}\cos{\omega_*}}}
\newcommand{\esinw}{\ensuremath{\sqrt{e}\sin{\omega_*}}}

\begin{document}

\title{Near-IR Emission Spectrum of WASP-103\MakeLowercase{b} \rm using {\it Hubble Space Telescope}/Wide Field Camera 3}

\author{Kimberly~M.~S.~Cartier\altaffilmark{1,2}, 
		{Thomas~G.~Beatty}\altaffilmark{1,2},
		{Ming~Zhao}\altaffilmark{1,2}, 
		{Michael~Line}\altaffilmark{3,4},
		{Henry~Ngo}\altaffilmark{3},
		{Dimitri~Mawet}\altaffilmark{9,10},
		{Keivan~G.~Stassun}\altaffilmark{5,6}, 
		{Jason~T.~Wright\altaffilmark{1,2}},
		{Laura~Kreidberg}\altaffilmark{7},
		{Jonathan~Fortney}\altaffilmark{8}		
		{Heather~Knutson}\altaffilmark{3},
}

\altaffiltext{1}{Department of Astronomy \& Astrophysics, The Pennsylvania State University, 525 Davey Lab, University Park, PA 16802, USA}
\altaffiltext{2}{Center for Exoplanets and Habitable Worlds, The Pennsylvania State University, University Park, PA 16802,, USA}
\altaffiltext{3}{Division of Geological and Planetary Sciences, California Institute of Technology, Pasadena, CA 91125, USA}
\altaffiltext{4}{NASA Ames Research Center, Space Science and Astrobiology Division, Moffett Field, CA 94035, USA}
\altaffiltext{5}{Department of Physics and Astronomy, Vanderbilt University, 6301 Stevenson Center, Nashville, TN 37235, USA
}
\altaffiltext{6}{Department of Physics, Fisk University, 1000 17th Avenue North, Nashville, TN 37208, USA}
\altaffiltext{7}{Department of Astronomy and Astrophysics, University of Chicago, 5640 S Ellis Ave, Chicago, IL 60637, USA}
\altaffiltext{8}{Department of Astronomy and Astrophysics, University of California, Santa Cruz, CA 95064, USA}
\altaffiltext{9}{Department of Astronomy, California Institute of Technology,Pasadena, CA, USA}
\altaffiltext{10}{Jet Propulsion Laboratory, California Institute of Technology,Pasadena, CA, USA}
\email{kms648@psu.edu} 
\footnotetext[]{Based on observations with the NASA/ESA {\it Hubble Space Telescope}, obtained at the Space Telescope Science Institute, which is operated by AURA, Inc., under NASA contract NAS 5-26555.}

\shortauthors{Cartier~et~al.}
\shorttitle{Near-Infrared Emission Emission Spectrum of WASP-103b}

\slugcomment{Accepted to AJ - Nov. 25, 2016}


\begin{abstract}
We present here our observations and analysis of the dayside emission spectrum of the hot Jupiter \planet. We observed \planet\ during secondary eclipse using two visits of the {\it Hubble Space Telescope} with the G141 grism on Wide Field Camera 3 in spatial scan mode. We generated secondary eclipse light curves of the planet in both blended white-light and spectrally binned wavechannels from $1.1-1.7 \micron$ and corrected the light curves for flux contamination from a nearby companion star. We modeled the detector systematics and secondary eclipse spectrum using Gaussian process regression and found that the near-IR emission spectrum of \planet\, is featureless across the observed near-IR region to down to a sensitivity of 175 ppm, and shows a shallow slope toward the red. The atmosphere has a single brightness temperature of ${\rm T_B=2890}$ K across this wavelength range. {This region of the spectrum is indistinguishable from isothermal, but may not manifest from a physically isothermal system, i.e. pseudo-isothermal.} A solar-metallicity profile with a thermal inversion layer at $10^{-2}$ bar fits the spectrum of \planet\ with high confidence, as do an isothermal profile with solar metallicity and a monotonically decreasing atmosphere with C/O\textgreater1. The data rule out a monotonically decreasing atmospheric profile with solar composition, and we rule out a low-metallicity decreasing profile as unphysical for this system. The pseudo-isothermal profile could be explained by a thermal inversion layer just above the layer probed by our observations, or by clouds or haze in the upper atmosphere. Transmission spectra at optical wavelengths would allow us to better distinguish between potential atmospheric models. 

\end{abstract}
\section{Introduction}
\label{sec:intro}
\setcounter{footnote}{0}


Spectroscopic measurements of exo-atmospheres are essential for a full characterization of exoplanet composition, temperature, and, eventually, habitability. Given the state of our current technology, transiting hot Jupiters, especially very hot Jupiters and ultra-short period Jupiters, are the best candidates for both transmission and emission spectroscopy because of their large radii, extended atmospheres, and hot equilibrium temperatures {\citep{char02,knutson07,snel08,deming13,kat16,sing16}}. Consequently, the study of exo-atmospheres has been largely limited to hot Jupiters, with super-Earths 55 Cancri e \citep{55cancrie}, HD 97658 \citep{knutson14b}, and GJ1214b \citep{bean10,desert11,berta12,kreidberg14} as notable exceptions with measured transmission spectra. Thermal emission spectroscopy, however, which measures the ratio of dayside planetary emission relative to the host star during secondary eclipse, is easily applied only to the hottest planets. By measuring the planet/star flux ratio as a function of wavelength, we can probe the atmospheric temperature at a range of pressures and heights to determine the vertical thermal profile of the atmosphere, and potentially detect the presence of molecular absorption.


A key feature of the Earth's atmospheric profile, the stratospheric temperature inversion, is caused by absorption of UV insulation by ozone, which is an essential atmospheric component for the protection of life. A similar temperature inversion in an exo-atmosphere, detectable by thermal emission spectroscopy, would be indicative of an analogous protective compound, and is therefore a highly sought-after atmospheric feature. While hot Jupiters are much too hot for life as we know it regardless of a temperature inversion, exo-atmospheric spectra have been consistently tested against atmospheric models containing temperature inversions to seek proof of concept for future application to cooler planets. As a result of temperature constraints, titanium oxide (TiO) or vanadium oxide (VO) are prime suspects for the additional heating of hot Jupiter stratospheres, rather than ozone or hydrocarbons \citep{fortney08,molliere15}.

However, evidence {against} a strong thermal inversion layer has been found for most exoplanets \citep{char08,mad10,mad11,brogi12,diamond14,stevenson14,schwarz15,line16}, supporting the hypothesis that inversions are only present in very highly irradiated hot Jupiter atmospheres ($\gtrsim2000{\rm\,K}$; \citealt{char08, fortney08}). \cite{spiegel09} suggests this may be due to titanium and vanadium being constrained to solids and raining out in all but the hottest atmospheres, which would require an unusually large amount of macroscopic mixing to overcome this and produce inversions. \cite{knutson10} postulate that the existence of temperature inversions might be limited by the incoming stellar UV flux that likely destroys TiO and VO in the exo-atmosphere. \cite{wasp33} and \cite{wasp33confirm} presented compelling evidence for the presence of a thermal inversion layer in the atmosphere of the highly irradiated hot Jupiter WASP-33b (${\rm T_{eq}=3000\,K}$). However, WASP-33b orbits a $7430$ K star, receives a large amount of stellar UV flux, and therefore challenges the theory put forth by \cite{knutson10}. Together, the hypotheses of \cite{spiegel09} and \cite{knutson10} suggest that thermal inversions will only be detectable in highly irradiated exo-atmospheres that receive low-UV flux, or have some mechanism to overcome TiO depletion.

The hot Jupiter \planet\ \citep{gillon14} is one of the best candidates for emission spectroscopy known to date. \planet\, has an orbital period of only 0.92 day and orbits at only $2.978$ times the stellar radius. This makes \planet\, one of the hottest known exoplanets with a zero-albedo, complete redistribution equilibrium temperature of $2890$ K. While being both highly irradiated and having an ultra-short period make \planet\, an ideal candidate for thermal emission spectroscopy, it also orbits a relatively quiet F8V star ($T_{\rm eff}=6110$ K) and receives low-UV flux compared to other ultra-short period hot Jupiters. This would allow us to test the the theory of \cite{knutson10} regarding the connection between incident UV flux and inversion strength by comparing two very hot planets (WASP-103b and WASP-33b) that receive different UV flux.

We have used Gaussian process (GP) regression to extract the first thermal emission spectrum of \planet\ from {\it Hubble Space Telescope}/Wide Field Camera 3 ({\it HST}/WFC3) observations of \planet\ at secondary eclipse. Gaussian process regression has previously been used to construct the transmission spectra of HD 189733b \citep{gibson12}, WASP-29b \citep{gibson13}, HAT-P-32b \citep{hatp32}, and most recently, CoRoT-1b \citep{corot1b}, and has been applied to eclipse observations of HD 209458b  \citep{hd209gp} and LHS 6343 C \citep{montet16} taken with {\it Spitzer}/Infrared Array Camera. This paper presents the first Gaussian process regression analysis of an exoplanet thermal emission spectrum taken with {\it HST}/WFC3.


We describe our {\it HST}/WFC3 observations, data calibration, and spectral extraction methods in Sec. \ref{sec:obs}. Sec. \ref{sec:second} details the detection of a nearby stellar source in our field of view, a probabilistic determination that the source is physically associated with \wasp, and the modeling of the spectral energy distribution of this source. Sec. \ref{sec:gp} outlines our Gaussian process regression method as applied to the blended white-light and the spectrally resolved eclipse light curves, compares our GP regression to a more traditional parametric regression technique, and presents the measured thermal emission spectrum of \planet. We present atmospheric modeling of that spectrum in Sec. \ref{sec:atmos} and compare the spectrum of \planet\ to those of other exoplanets to supplement our interpretation the atmospheric profile of \planet\ and motivate future studies.
\section{Secondary Eclipse Observations}
\label{sec:obs}

\begin{table}
\begin{center}
\caption{Summary of Spatial Scan Observations}
\label{tab:obs}
\begin{tabular}{lcc}
\hline
										&Visit 1	&	Visit 2 \\ \hline\hline
Date of observations (UT)	&2015 Jun 15 & 2015 Jun 17\\
Time of first scan (JD)		&2457189.2036& 2457191.0608\\
Time of last scan (JD)			&2457189.4979& 2457191.3540\\
Number of {\it HST} orbits	&5&5\\
Observations per orbit\tablenotemark{*}		&(11)12&(11)12\\
Total number of observations&118&118\\
Scan rate (\arcsec/s)			&0.025&0.025\\
Scan duration (s)				&81.089&81.089\\
Detector subarray size (pixels)&$256\times256$&$256\times256$\\
Median S/N per spectral column&{5946.43}&{5871.51}\\ \hline
\end{tabular}
\end{center}
\vspace{-0.3cm}
\tablenotemark{*}{The first orbit in each visit had only 11 observations, while orbits 2, 3, 4, and 5 in each contained 12 observations.}

\end{table}

We observed \wasp\, during two visits of {\it HST} on UT 2015 June 15 and 2015 June 17, and used the WFC3-IR camera and the G141 grism in spatial scan mode to provide slitless spectroscopy at wavelengths from $1.1\micron$ to $1.7\micron$. Details of our observations are found in Table \ref{tab:obs}. We obtained 10 orbits in total over the two visits, with scan durations of $81.089$ s using SPARS10 and ${\rm NSAMP=12}$. This multivisit approach has been used in many recent WFC3-IR G141 observations of transiting planets, which have generated reliable high-precision results (e.g., 15 visits in \cite{kreidberg14}; 4 visits in \cite{knutson14a}; 2 visits in \cite{huitson13}). The second visit of the eclipse, which proved to be consistent with the first, demonstrates repeatability and allows us to achieve higher precision at similar spectral resolution (see Sec. \ref{sec:tes}).

As shown in previous observations \citep{berta12,deming13,knutson14a,kreidberg14}, the first orbit of a new WFC3-IR observing sequence always displays larger-than-usual instrumental effects as the charge traps fill from an empty state before reaching steady state in subsequent orbits \citep{wfc3ir}. We used the first orbit of each visit to capture these instrument systematics and used orbits 2 through 5 to observe the eclipse and pre- and post-eclipse baselines. Use of spatial scan mode increased the observing efficiency, minimized detector systematics caused by imperfect flat fielding, and allowed for longer observing times without saturation. We alternated between forward and reverse scan directions in order to further reduce overheads, as previously demonstrated in \citet{kreidberg14} and \citet{knutson14b}.

We used the $256\times256$ pixel subarray mode to reduce both readout time and data volume, which minimized overhead and time loss due to serial buffer dumps.
{The signal-to-noise ratio (S/N) per spectral column of each visit spanned an order of magnitude across the dispersion direction, peaking near the middle of the wavelength range, and falling off toward the edges. For Visit 1, the S/N ranged from ${\rm S/N=32.14-7376.23}$, with a median value of $5946.43$, and the Visit 2 S/N ranged from ${\rm S/N=26.37-7371.68}$, with a median value of 5871.51 (Table \ref{tab:obs}).} When measured across several (6-8) binned columns, this yielded an average regression scatter of 515 ppm for Visit 1 and 543 ppm for Visit 2, which are 3.7 and 3.9 times the photon noise levels of 139 ppm and 138 ppm, respectively. {Read noise for the WFC3/IR detector is between 10 and 20 electrons according to \cite{wfcihb}. This corresponds to a read noise of 0.4 ppm for a white-light curve and 7.4 ppm for a spectrally resolved light curve near the middle of the detector.}

\subsection{Background Subtraction, Flat Fielding, Subframe Alignment, Cosmic Ray Correction}
\label{sec:calib}

{To better subtract background and account for a tiny dispersion drift during spatial scan, we subtracted sequential pairs of up-the-ramp readouts within each exposure ($81.089$ s) to generate a set of subframe images. Each subframe image represents a shorter exposure of $7.347$ s along the spatial scan direction.}

 

{Because our scan speed is very slow ($0\farcs025 {\rm s}^{-1}$), the point spread function (PSF) of each subframe image is substantially undersampled. This resulted in varying subframe PSFs that could not be combined as an average for outlier rejection because of under sampling and changing centroid while scanning.} Therefore, we used the subframe only for trimming nearby contaminations, removing background (due to the trimming), and realignment in wavelength.

After examining the subframes, we found that the starlight from \wasp\, is strictly constrained in a relatively small area, outside of which there is only background flux. We defined a conservative mask for {subframe images} and used the remainder of the readout outside the mask to determine and subtract the background from {each subframe image. The background is spatially flat and uniform due to the short exposure time.} That background was subtracted so that all pixels in the background area {of each subframe image} would be zero plus noise. {We then defined a smaller mask following \cite{deming13} and \cite{knutson14a}, and zeroed all pixels outside of the mask. This helps to reduce noise and exclude cosmic rays (CRs)  in the background area when later combining all subframes to determine the flux for each exposure.} We found the optimal trim height (from the center of the image ``band'') to be 25 pixels, which excludes 1.12\% of total flux from the extended halo of the PSF per image.

As discussed in \cite{mandell13}, the {\tt .flt} images provided by the WFC3 {\tt calwf3}\footnote{Version 3.3; \url{http://www.stsci.edu/hst/wfc3/pipeline/wfc3\_pipeline}} calibration pipeline often yield time series with higher RMS than those with flats  produced by {\tt .ima} files. We therefore chose to create our own flat fields for data reduction. We determined the centroid in both the dispersion direction ($X$) and scan direction ($Y$) of each subframe and checked if there was any significant drift in the $X-$position of the scan on the detector. We found that the $X-$position drifts from different up-the-ramp subframes were well within $\sim0.05$ pixels. {To examine if this tiny drift can affect final extracted flux, we convolved each column-summed subframe spectrum with a 5-pixel Gaussian kernel and then cross-correlated and aligned each subframe using a cubic spline interpolation. The resulting summed fluxes for each exposure agree well with each other before and after the alignment.} Any added uncertainty from these shifts were negated by wavechannel binning (see Sec. \ref{sec:spec}).

We used the centroid information and the initial image position to generate flat fields, assuming each column had the same wavelength{, since the column direction is perpendicular to the dispersion direction}. We applied these flat fields to each subframe. The dispersion drift along each column was small and was accounted for during {subframe alignment}, and produced negligible effects during wavelength calibration.

We identified and corrected for {additional} CRs and bad pixels {on the PSF that were not trimmed by the mask} by {first} taking the average of multiple exposures (not subframes) from each orbit and same scan direction for normalization. We then applied a moving median filter to reject additional bad pixels and CR hits {for the normalized image}. Two applications of the filter with slightly different median rejection windows removed all visible spurious effects. Because of the slight drift between two different scan directions, each direction needed to be treated separately.


\subsection{Spectral Extraction}
\label{sec:spec}

\begin{figure}[t]
	\begin{center}
	\includegraphics[width=0.45\textwidth]{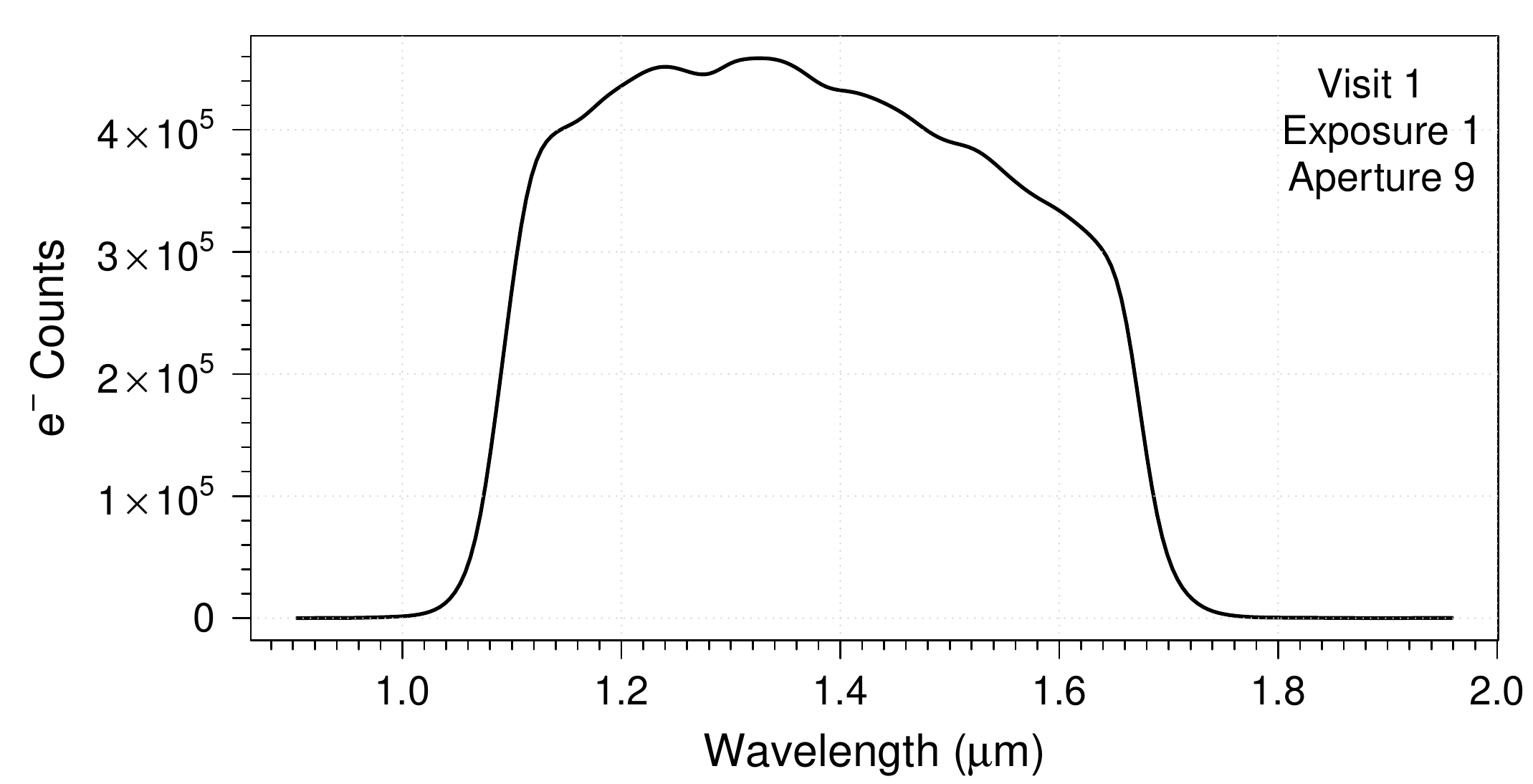}
	\caption{A representative spectrum for a single exposure after summing along each pixel column, with both scan directions combined.}
	\label{fig:expspec}
	\end{center}
\end{figure}

\begin{figure}[t]
	\begin{center}
	\includegraphics[width=0.45\textwidth]{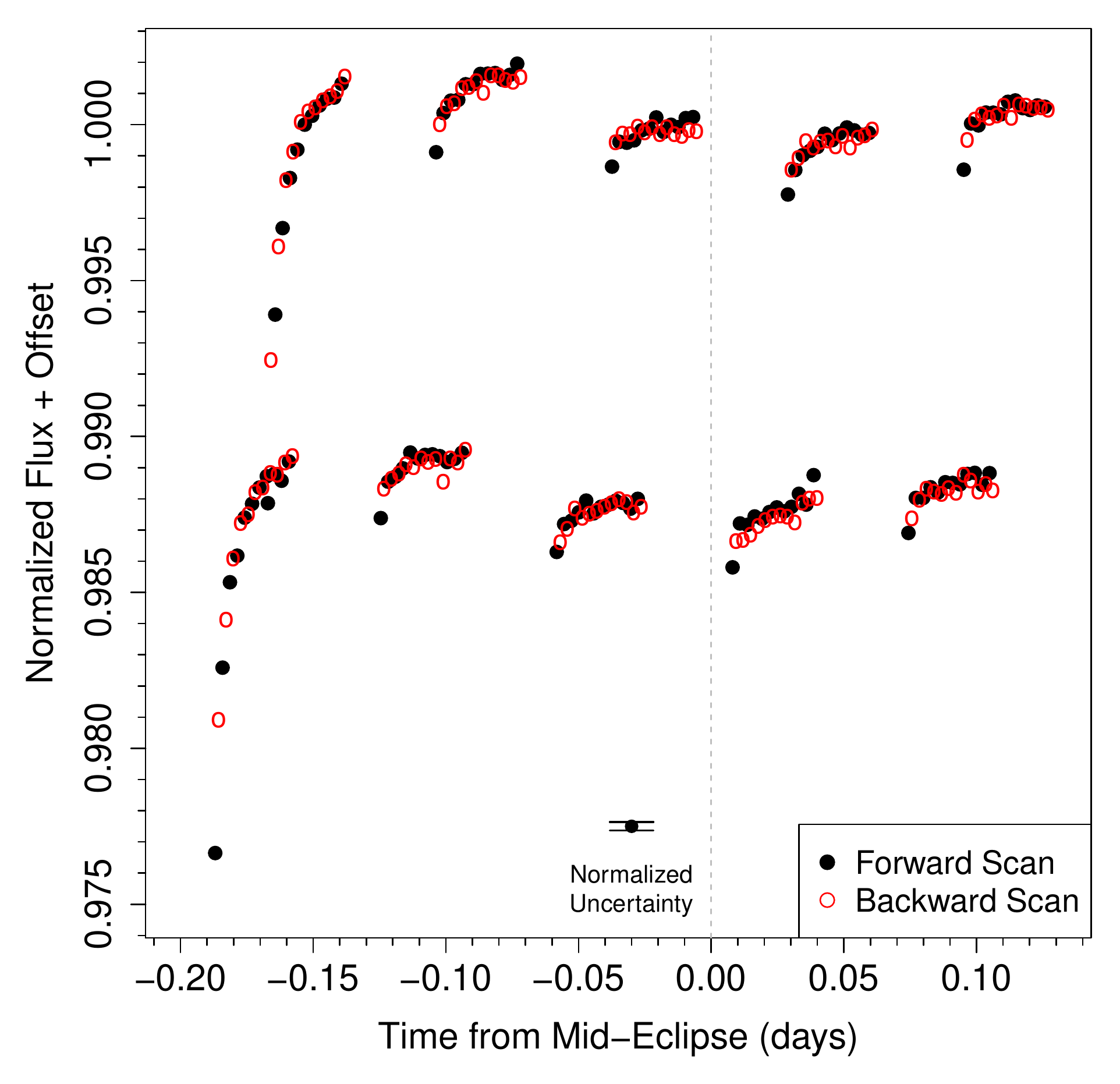}
	\caption{Normalized white-light flux time series for Visit 1 and Visit 2, measured in time from mid-eclipse. Solid black points represent the time series in the forward-scan direction, and open red points represent the backward-scan direction. The vertical offset has been added for clarity. Vertical dashed gray line indicates mid-eclipse. A representative error bar of normalized uncertainty in flux values is included for reference.}
	\label{fig:raw}
	\end{center}
\end{figure}

We used the background-subtracted CR-corrected scans to extract the ``white-light'' flux time series in both the forward- and backward-scan directions. We summed along each column of the subframe-aligned exposures to construct a single spectrum for each exposure (Fig. \ref{fig:expspec}), which we then used to convolve, interpolate, and calibrate each exposure to the same wavelength scale. The spectrum for each exposure was then integrated across wavelength to generate a single white-light flux for each exposure, and it was used to create the white-light time series shown in Fig. \ref{fig:raw}. We used the photon noise of each exposure, calculated as $1/\sqrt{raw\,flux}$, as the uncertainty for each flux point in the time series. We found that any sources of noise unaccounted for in this uncertainty were later accounted for in the GP regression (see Sec. \ref{sec:gp}). We found the average uncertainty of the white-light time series to be $\sigma_{\rm flux}=138$ ppm in normalized flux units{, well above the read noise of $\sim few$ ppm.}

In spatial scan mode, light from the target is dispersed by the grism onto the detector, which means that the wavelength solution of each exposure is sensitive to the $X-Y$ position of the image on the detector. To create the spectrally resolved time series for each visit and scan direction, we determined the $X-$ and $Y-$centroids of each exposure, aligned each extraction aperture based on the centroids, and extracted the flux using partial pixels. This alignment accounted for any centroid shifts between exposures. We then summed the aligned spectra for incremental aperture sizes in $Y$ and $X$ directions. This method allowed for later optimization of the apertures for spectrophotometric extraction. The aperture optimization is discussed in more detail in Sec. \ref{sec:aperture}. 

The spectrum of each exposure was binned into 22 wavechannels spanning the first-order wavelength range of the grism, at a constant $\Delta\lambda= 0.02788\micron$ for all but the edge bins, which were slightly wider in wavelength. The spectrally resolved time series show similar systematics to those seen in Fig. \ref{fig:raw}. We found that the common centroid of the aligned spectra in the dispersion direction differed from Visit 1 to Visit 2 by $\Delta {\rm X_ {V1-V2}}=-0.105$px, which produced slightly different wavelength solutions for each visit. We found that the shortest wavelength from Visit 1 was $1.3\times10^{-6}\micron$ shorter than for Visit 2, and that the Visit 1 spectrum covered a wavelength range $4\times10^{-7}\micron$ wider than Visit 2. When binned into the spectral wavechannels, this caused a shift in central bin wavelength of $\Delta\lambda_{V1-V2}=-0.00465\,\micron$ for all wavechannels. This slight shift in central wavelength was accounted for in the flux decontamination of a nearby star (Sec. \ref{sec:second}) and construction of the visit-averaged thermal emission spectrum of \planet\ (Sec. \ref{sec:tes}).


\subsection{Aperture Optimization}
\label{sec:aperture}

\begin{figure*}[t]
	\begin{center}
	\includegraphics[width=0.9\textwidth]{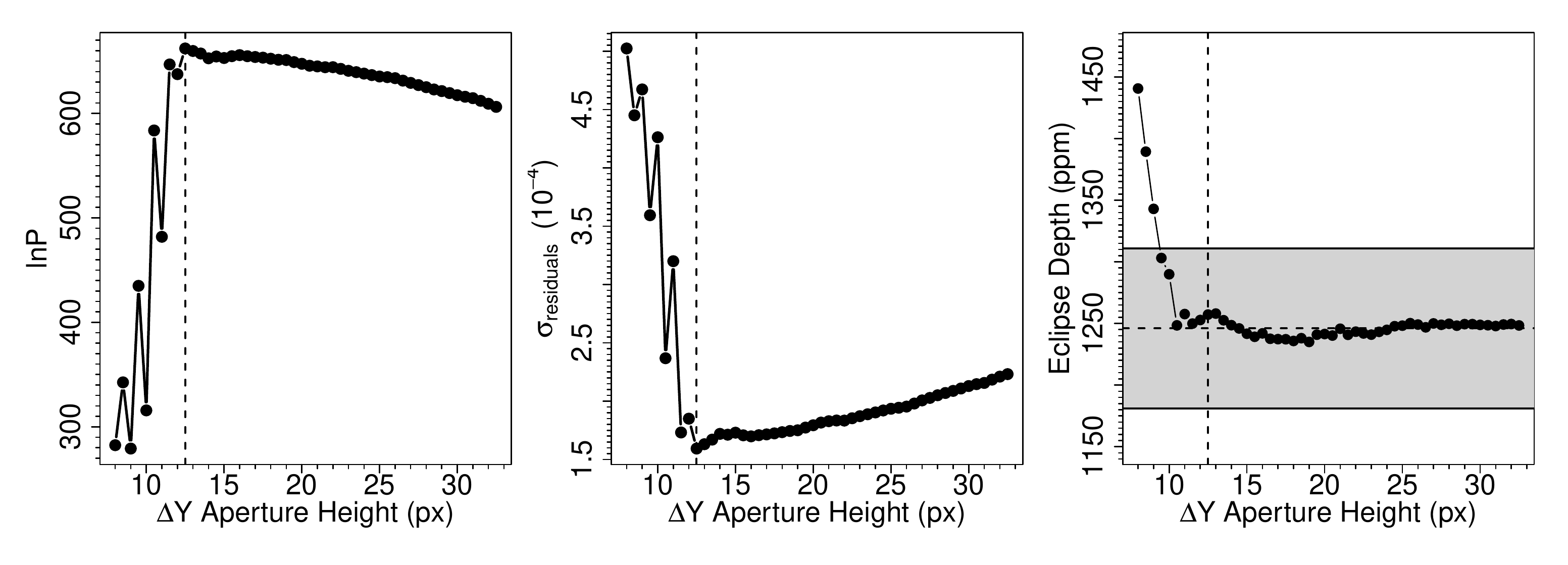}
	\includegraphics[width=0.9\textwidth]{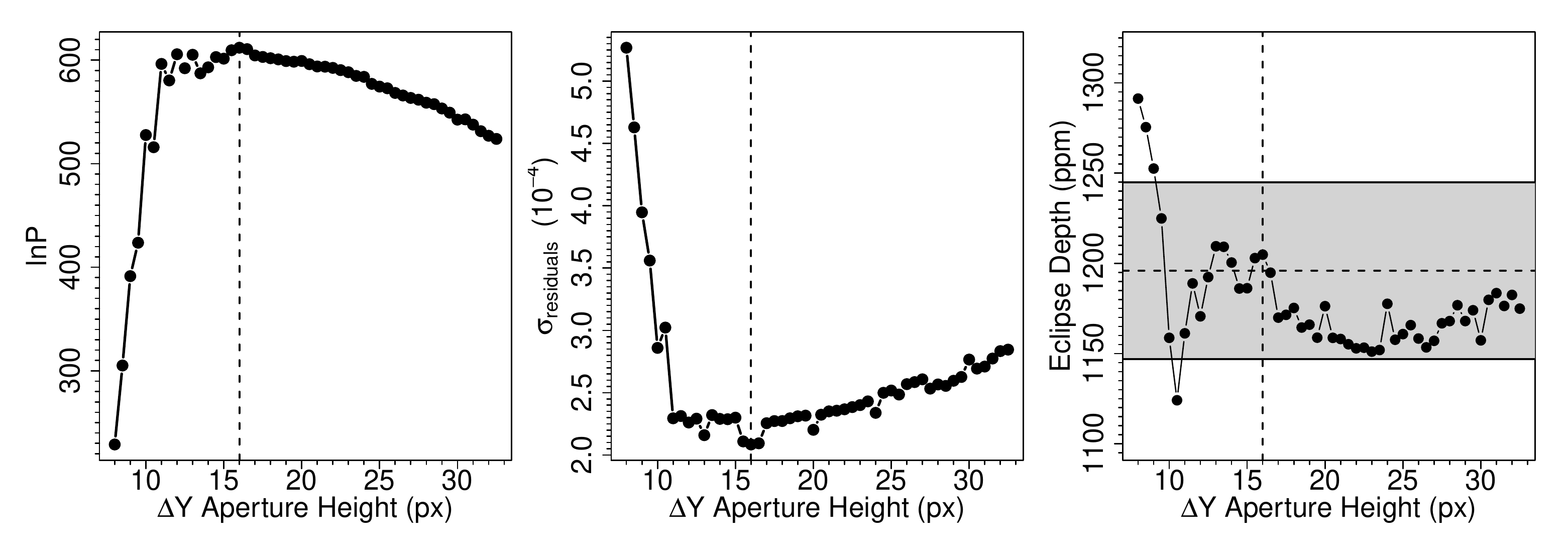}
	\caption{Aperture optimization for Visit 1 (top row) and Visit 2 (bottom row), showing changing $\ln P$ (left), standard deviation of the residuals (center), and eclipse depth (right) as a function of aperture height from $Y-$centroid. The vertical dashed line indicates the optimal aperture. In the rightmost panel, the horizontal dashed line and shaded region indicates the white-light eclipse depth and $1\sigma$ uncertainties at the optimum aperture fitted with the MCMC sampling.}
	\label{fig:ap}
	\end{center}
\end{figure*}


We found the optimum aperture for spectrophotometric extraction of each scan to determine how much the extraction aperture affected our results. We first tested for the optimum aperture in the dispersion ($X$) direction by extracting white-light spectra at 60 $\Delta{X}$ apertures starting at ${X_{\rm centroid}}\pm 60$ pixels and increasing in 0.5-pixel increments, keeping the aperture in the scan direction fixed. We chose the aperture that produced the lowest BIC at a fixed $\Delta{Y}$, and found that a dispersion aperture of $\Delta{X}=68$ pixels from the $X-$centroid is preferred.

We then tested for the optimum aperture in the scan ($Y$) direction by extracting white-light spectra at 50 $\Delta{Y}$ aperture heights starting at ${Y_{\rm centroid}}\pm 8$pixels and increasing in 0.5 pixel increments, keeping the $\Delta{X}$ aperture fixed at its optimum value. We then applied the initial maximization procedure as detailed in Sec. \ref{sec:regression} to the white-light eclipse curves extracted at these apertures, keeping fixed all parameters describing the physical and orbital condition of \planet\, (i.e. system parameters as defined in Sec. \ref{sec:gp}) and only fitting for parameters associated with detector systematics, eclipse depth, and eclipse time (i.e. hyper- and eclipse parameters as per Sec. \ref{sec:gp}).  This was done for both visits using only the initial {\tt amoeba} maximization (downhill simplex method; \citealt{amoeba}). We did not correct for contamination from the companion star described in Sec. \ref{sec:second}. We chose the aperture with the maximum log-likelihood (Eqn. \ref{eq:lnp}), lowest residual scatter, and a stable eclipse depth as the optimum $\Delta Y$ aperture for spectrophotometry. 

Fig. \ref{fig:ap} shows the results of the aperture optimization for Visit 1 (top) and Visit 2 (bottom). We find that $\Delta{Y}=12.5$ pixels is optimum for Visit 1 and $\Delta{Y}=16.0$ pixels is optimum for Visit 2. The scatter in eclipse depth across apertures after the eclipse depth has stabilized is 5.8 ppm and 16.5 ppm for Visit 1 and Visit 2, respectively, which are much smaller than our final eclipse depth uncertainties of 63 ppm and 49 ppm. We are confident that a slight deviation in the height of our chosen aperture would not significantly impact our results. 

\section{Detection of the Companion Star}
\label{sec:second}

\begin{figure}[t]
	\label{fig:nirc2}
	\includegraphics[width=0.45\textwidth]{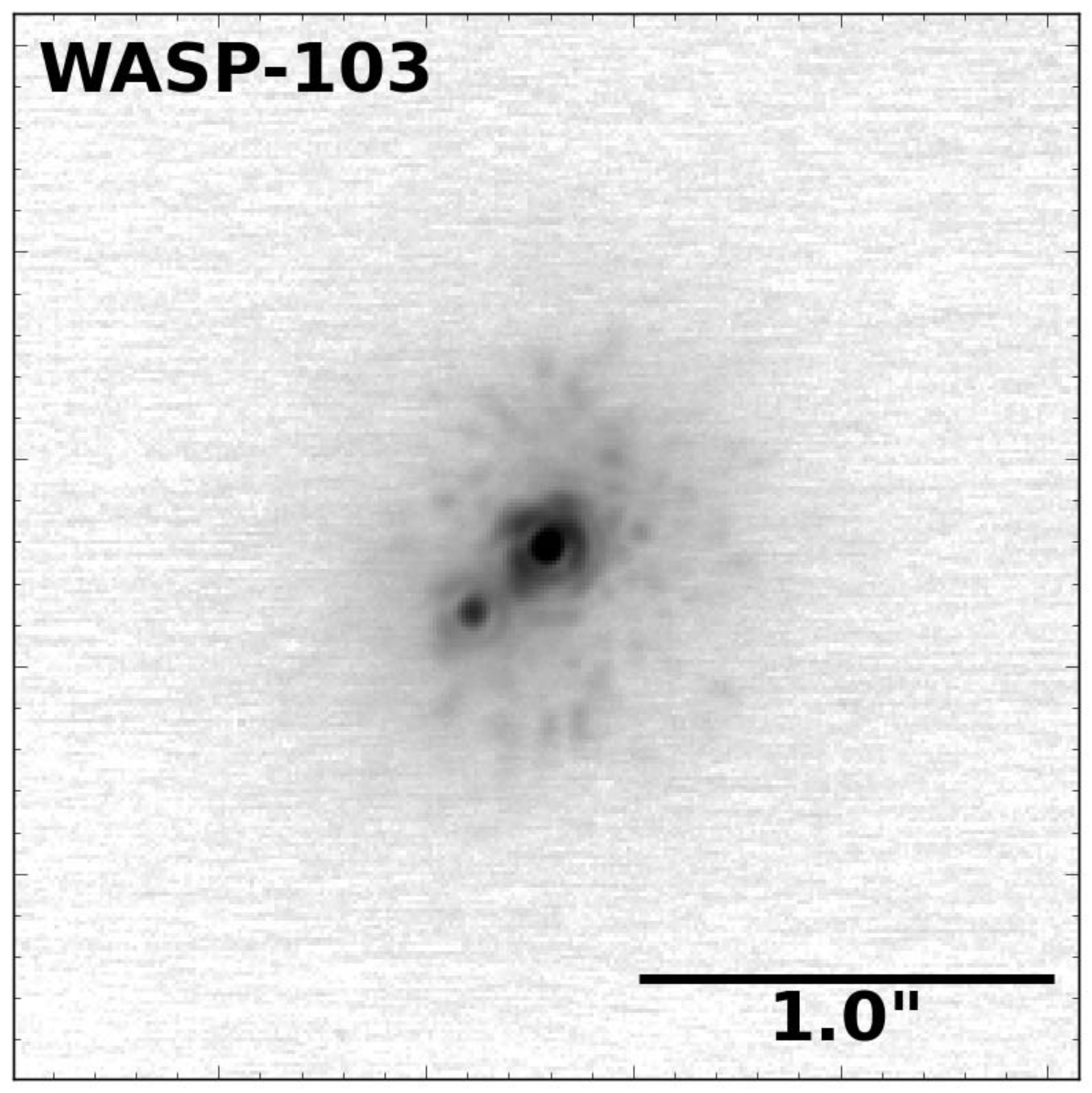}
	\caption{Keck NIRC2 AO image of the \wasp\, system in the $K_S$ band, with $1\farcs0$ marked for scale. In this image, the upper right star is the primary (A), and the bottom left star is the companion star (B). North is oriented upwards and east is oriented to the left. Intensity is on a logarithmic scale.}
\end{figure}

\begin{figure*}[t]
	\begin{center}
	\includegraphics[width=\textwidth]{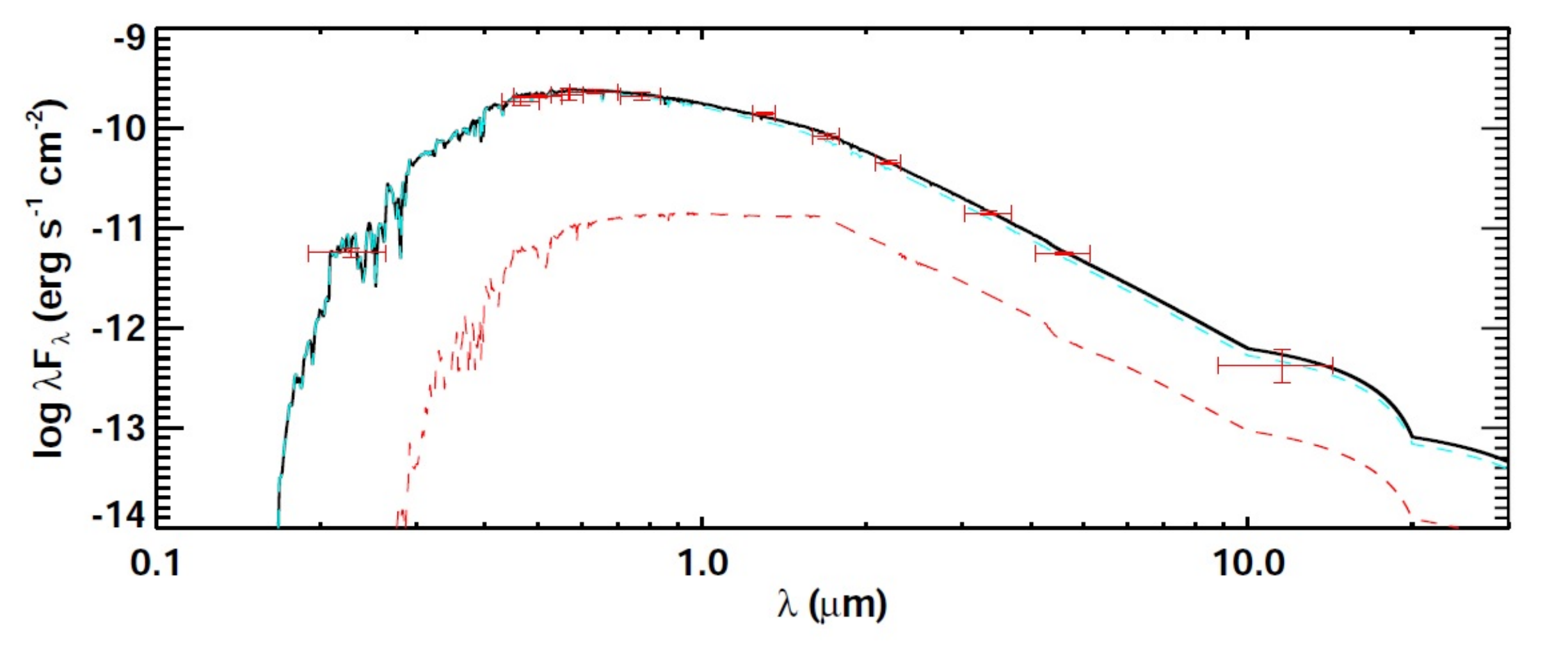}
	\caption{SED fits for primary and companion stars. The cyan dashed line is the primary star (A), the red dashed line is the companion star (B), and the solid black line is the reblended A+B SED. Red crosshairs show the observed blended photometry, with horizontal error bars representing the width of the passbands.}
	\label{fig:sed}
	\end{center}
\end{figure*}

\begin{table}[t]
\begin{center}
\caption{Photometry of the Companion Star}
\label{tab:companion}
\begin{tabular}{cccc}

\hline\hline
	 								&	$J$ band					&	$H$ band 					&	$K_S$ band \\ \hline
Blended Photo					&	$11.100\pm0.023$	&	$10.857\pm0.030$ 		&	$10.767\pm0.018$ \\
-metry (mag)&&&\\
$\Delta ({\rm B-A})$ (mag)	&	$2.45\pm0.07$	 		&	$2.21\pm0.05$ 			&	$2.06\pm0.03$ \\
Separation (mas)				&	$240.5\pm 1.5$ 		&	$239.8 \pm 1.4$ 			&	$239.7 \pm 1.5$ \\
Position Angle					&	$131\fdg36\pm0\fdg35$ &$131\fdg38\pm0\fdg35$&$131\fdg41\pm0\fdg35$ \\ \hline 
\end{tabular}
\end{center}
\vspace{-0.3cm}
\tablecomments{Blended photometry is from 2MASS. Separation and position angle are from \cite{ngo16}, and $\Delta{\rm (B-A)}$ is from our own PSF-fitting method.}
\end{table}

\cite{lucky15} reported the detection of a previously unknown stellar source $0\farcs242\pm0\farcs016$ away from \wasp\, in $i'$ and $z'$. We imaged the \wasp\, system again in 2016 January and confirmed the nearby source using Keck NIRC2 AO observations in $JHK_S$. Fig. \ref{fig:nirc2} shows a $2\farcs5\times2\farcs5$ snapshot of the full $10\arcsec\times10\arcsec$ image in $K_S$. There were no additional stars observed in the full NIRC2 image.

We reduced the NIRC2 images and calculated the photometry of the companion following an approach similar to that used in \cite{zhao14} and \cite{bechter14}. We measured the flux ratios of \wasp\, and the companion star in $J$, $H$, and $K_S$ bands by simultaneously fitting PSF models for both stars. We used a Gaussian function to characterize the core of the PSF and a Moffat function to trace the extended PSF halo. Because of high Strehl ratios in the $H$ and $K_S$ bands, the images are diffraction limited and Airy rings were clearly seen in these two bands. We therefore added an Airy function to the PSF model to account for the diffraction pattern. To examine the effects of different PSF components and avoid overfitting, we fit each image with three sets of models: {1. sum of Moffat and Gaussian}; 2. sum of Airy and Gaussian; 3. sum of Airy, Gaussian, and Moffat\footnote{Ideally, an obscured Airy function should be used. However, the low S/N ratio of the Airy rings in the images means that a normal Airy function with a Gaussian component in the center can still properly model the bulk of the ring patterns.}. We selected the best model using the Bayesian information criterion (BIC).
We assumed the same PSF shape for the two stars and only allowed their flux ratio to vary. Because of the high Strehl ratios in $H$ and $K_S$, the Airy, Gaussian, and Moffat is preferred in these two bands, while the Gaussian and Moffat model is preferred in the $J$ band where the Airy rings are overwhelmed by the PSF halo. 

To better model the extended PSF halo while avoid fitting on the noisy sky background, we limited the fitting range to two circular apertures of the same size (see below), centered on the centroids of the two stars.  Because of their proximity, the flux of the companion star depends on the halo of \wasp, which in turn depends on the size of the aperture. To avoid bias in choosing the best size for the field of view, which is hard to determine becuase of the noisy background and the faintness of the PSF halo, we fit the PSFs with a set of aperture sizes ranging from 10 to 30 pixels with a step size of 1 pixel. The flux ratio stabilizes beyond 20 pixels once the halo S/N is low and becomes dominated by background noise. The final value and uncertainty of the flux ratios are determined by taking the median and standard deviation resulting from all aperture sizes and all images in each band\footnote{Traditional model selection techniques such as BIC are not suitable here because changing the aperture size also changes the data in the fit since many more background pixels will be included (proportional to radius$^2$). Thus, comparing BICs means comparing different data sets. As a result, the minimization of least-square residual will be biased toward fitting the background rather than the PSF cores in the center. To avoid that bias and bias caused by using a small aperture, we take the median (instead of the mean for robustness) and scatter of all aperture fits to conservatively account for the variation in flux ratio caused by imperfect modeling of the overall PSF.} .

Relative to \wasp, we find that the companion star has a photometry ${\Delta J = 2.45\pm0.07}$, ${\Delta H= 2.21\pm0.05}$, and ${\Delta K_S = 2.06\pm0.03}$ (Table \ref{tab:companion}). The NIRC2 AO observations are also published in \cite{ngo16}. Their reduction and PSF-fitting method of the candidate companion yielded photometry consistent with a K or M spectral type, and is described in detail in \cite{ngo15,ngo16}. Our reduction and PSF methods yielded photometry values consistent with those provided in \cite{ngo16}. From the analysis of \cite{ngo16}, the separation between \wasp\ and the companion star is $240.0\pm1.5$ milliarcseconds when averaged between bandpasses, and the companion star is located at an average position angle of $131\fdg37\pm0\fdg35$.

The companion star (source B) contributes a significant amount of flux in the near-infrared (NIR), and has a small enough separation from the \wasp\ (source A) that it contaminates our observed secondary eclipse light curves. In order to estimate the flux contamination from the companion star as a function of wavelength, we determined the spectral energy distribution (SED) of the companion star. The flux contamination ratio, $F_B/F_A$, is primarily only dependent on the effective temperature of the two stars. The SEDs of a dwarf star and a giant star at the same effective temperatures do not vary significantly at the wavelengths of interest, and so it does not have significant bearing on our SED fit whether \wasp\ and the companion star are physically associated. However, we first determined probabilistically whether the companion star is likely located at the same distance as \wasp, then modeled the SED of the companion. 

Based on highly uncertain astrometric measurements of common proper motion and companion separation, \cite{ngo16} were unable to conclusively determine the physical association between \wasp\, and the companion star, but retained it as a ``candidate companion.'' The high galactic latitude of the \wasp\, system ($l=23.4099, b=+33.0215$) implies a low background stellar density and low probability of the random superposition of a background or foreground star. However, we quantitatively validated this assumption by simulating the stellar population along the line of sight for various fields of view (FoV) using the Besan\c{c}on stellar population synthesis of the Milky Way \citep{besancon}\footnote{\url{http://model.obs-besancon.fr/}}. The Besan\c{c}on online utility generates a list of stars that could theoretically be observed in a given FoV, with the option of restricting the generated list of stars based on luminosity class, galactic structure component, and a variety of stellar and observational parameters.

We wish to determine the probability of a star at least as bright as the companion star being observed in the NIRC2 images, and we wish to know the likelihood of any of those observed stars to be physically bound to \wasp. \cite{rag10} find that physically bound companion stars follow a Gaussian distribution with respect to $\log P$ measured in days, with $\mu_{\log P}=5.03$ and $\sigma_{\log P} = 2.28$. Statistically, the majority of companion stars with $\log P_C\leq(\mu_{\log P}+\sigma_{\log P})$ will be physically bound to the primary star. Assuming a total stellar mass of $1.5\msun$, this upper limit on $\log P$ is equal to a physical separation of $\sim10^3$ AU. When observed at the distance of the \wasp\ system ($d=470\pm35\ \pc$), this physical separation corresponds to a projected angular separation of $2\farcs13$. Therefore, we choose to simulate a circular FoV with a radius $r=2\farcs13$ centered on \wasp, and can then state that any stars that fall within this FoV are likely to be bound to \wasp.

Our ``bound star'' FoV with a radius $r=2\farcs13$ has an area $1.1\times10^{-6}{\rm deg^2}$, which is smaller than the resolution of the Besan\c{c}on simulation (minimum resolution of 0.01 deg$^2$). In order to calculate the probability that a single star with the observed photometry could randomly fall into the bound star FoV, we simulated 28 fields of view logarithmically spaced between $0.01{\rm\ deg^2}{\rm\, to\,}10{\rm\ deg^2}$ and applied the Poisson probability distribution to extrapolate the probability of a single star in our field of view. For each tested FoV, we simulated a full stellar population (i.e. no assumptions on luminosity class or galactic population). To determine the probability of detecting a star {at least as bright as the companion star}, we filtered out stars more than $3\sigma$ fainter than the companion star in $J$, $H$, and $K_S$ (Table \ref{tab:companion}). 

We fit a linear model to the resulting star counts as a function of FoV area ($\chi^2_{lin}/{\rm d.o.f.}=1.767$) and found that an average of $n=1677.733$ stars at least as bright as the companion star are expected for a 1 deg$^2$ FoV. We used the linear model to scale this value to find the average number of stars expected in our bound star FoV. Finally, using the Poisson probability distribution, we calculate ${P(1\ \text{star\ in\ the\ bound\ star\ FoV}) = 1.842\times10^{-3}}$. From this we conclude that the source B is likely not a random fore- or background star superimposed on the NIRC2 image, but instead is likely physically associated with the \wasp\, system.

Presuming the two stars are physically associated, we fit a theoretical SED to the blended flux from \wasp\ and the companion star using two dwarf star spectra with model atmospheres from \cite{kurucz}. {The \cite{kurucz} models with $\mathrm{T_{eff}} = 4000-7000$ K,$\log g = 2.0-4.5$ cgs, and$\mathrm{[Fe/H]} = -1.5-0.5$ dex were selected for fitting}.  All available photometry was used in the SED fitting, including GALEX near-UV, APASS $BVgri$, 2MASS $JHK_S$, and W1 to W3 from WISE. We required that the two SED components obey the $J$, $H$, and $K_S$ band $\Delta$-magnitudes from the NIRC2 image listed in Table \ref{tab:companion}, and applied prior stellar parameters about the primary star \citep{gillon14,southworth} to separate out the contributions to the combined flux from each star. The SED fitting also allowed ${\rm A_v}$ to be a fitted parameter, limited by the maximum line of sight ${\rm A_v}$ from the \cite{schlegel} dust maps. The theoretical SED solutions were verified by reblending them and determining the goodness-of-fit to the observed blended photometry.

The final reblended SED solution reproduced the observed blended photometry with a reduced $\chi^2 = 1.02$, and is shown in Fig. \ref{fig:sed}. The SED fit indicates  $T_{\rm eff} = 4400 \pm 200$K for the companion star, which is consistent with a K5 V spectral type \citep{kdwarf}. Using the bolometric flux ratio from the SED fits and the ${T_{\rm eff}}$ ratio, we obtain a radius ratio of ${\rm R_B / R_A} = 0.52\pm 0.05$. This SED solution is consistent with the values reported by \cite{ngo16} and \cite{newsouth}, who also report the mass of the companion star as $0.72\pm0.08\msun$. We used the SED fit here to later correct the secondary eclipse spectrum of \planet\ for contamination from the companion star. The flux decontamination is detailed in Sec. \ref{sec:contam}.

\section{Gaussian Process Regression of Light Curves}
\label{sec:gp}

Previous studies have successfully applied parametric models to capture detector systematics in spatial scan mode, and have been applied to emission and transmission spectroscopy for a multitude of exoplanets (\citealt{hd189733b,knutson14b}, for example). Enforcing a prespecified choice of parametric model of the systematics works well when the form of the systematics is known {\it a priori} or can be easily determined. Although we have functional forms for the WFC3 systematics in the form of a linear trend over a visit and an exponential ramp within an orbit \citep{deming13,knutson14b,corot2b}, and we have some indication that the ramp may be due to charge trapping in the detector \citep{agol10}, we lack a clear understanding why it would take that specific shape. Therefore, we choose a more flexible method for modeling the instrumental effects in our {\it HST}/WFC3 light curves.

GP regression eliminates the need for prespecifying a parametric model of the unknown systematics in favor of a more elastic representation of systematics and long-term trends (for a more in-depth discussion, see \citealt{gpbook,gibson12,kernel}). \cite{gibson11} was the first to demonstrate the necessity of GP regression for modeling {\it HST}/NICMOS systematics. Subsequently, the successful application of GP regression was demonstrated on the {\it HST}/NICMOS \citep{gibson12} and later {\it HST}/WFC3 \citep{gibson12b} transmission spectra of HD 189733b. While the uncertainties reported by a GP regression will often be larger than those reported by regression using a parametric model, these uncertainties and parameter values will likely be more accurate. 

To find the best-fit secondary eclipse model of our data via GP regression, we calculated the likelihood function, $L_{\rm model}$, given by

\begin{equation}
L(\mathbf{r}|\mathbf{X},\Phi)_{\rm model}=\frac{1}{ (2\pi)^{\sfrac{n}{2}} |\mathbf{\Sigma}|^{\sfrac{1}{2}} } \exp\left(-\frac{1}{2}\mathbf{r}^T\mathbf{\Sigma}^{-1}\mathbf{r} \right)
\label{eq:gplike}
\end{equation}

\noindent where $\mathbf{r}$ is the vector of residuals between the data values and the eclipse model, $\mathbf{X}$ is the vector of data locations (i.e. observation times), $\Phi$ is the set of {hyperparameters} that characterize the behavior of the covariance matrix $\mathbf{\Sigma}$, and $n$ is the number of data points. This likelihood function is a multivariate normal distribution. The secondary eclipse light-curve model \citep{mandel} is explicitly calculated as part of the residual vector and is the mean of the multivariate normal distribution. 

The covariance matrix, $\mathbf{\Sigma}$, captures the behavior of the data that cannot be attributed to the eclipse model and depicts how each value depends on each other value in the set. The matrix is populated by a covariance kernel, and by choosing an appropriate kernel to populate the covariance matrix, we can account for the effects of detector systematics without prespecifying a parametric model for these systematics. The residuals to the model should not exhibit any non-normal behavior if an appropriate kernel is chosen.

We observed in our data that not only are sequential points correlated with each other, but there is also periodicity in the correlation that corresponds to each {\it HST} orbit. This is easily seen in Fig. \ref{fig:raw} as a linear trend across the separate orbits and an exponential ramp creating a hook shape within each orbit. Therefore, we chose a quasi-periodic kernel to populate the individual elements of the covariance matrix \citep{kernel}. Each matrix element $\Sigma_{ij}$ is given by   

\begin{align}
\label{eq:covar}
\Sigma_{ij} = A^2 \exp \bigg(	&- \frac{\sin^2\left[\pi(x_i - x_j)/\theta\right]}{2\Omega^2} \\ \nonumber
						&- \frac{(x_i-x_j)^2}{L^2}\bigg)+\delta_{ij}\sigma_i^2 \nonumber
\end{align}

\noindent where $A$ is the amplitude of the covariance kernel, $\theta$ is the characteristic timescale of the periodicity, $\Omega$ is the coherence scale of the periodicity, $L$ is the characteristic time lag, $\delta_{ij}$ is the Kronecker delta, and $\sigma_i$ is the white-noise uncertainty associated with data point $x_i$. $A,\, $L$,\,\theta,$ and $\Omega$ are the four {\it hyperparameters} that characterize the covariance kernel ($\Phi$ in Eq. \ref{eq:gplike}) and capture the behavior of the instrument systematics. The periodic component of Eq. \ref{eq:covar} containing $\theta$ and $\Omega$ accounts for the exponential ramp, and the squared-exponential component containing $L$ accounts for the linear trend in the light curve.

We also included $\rp/\rs$, $\aplan/\rs$, $\cos i$, \esinw, \ecosw, $\rs$, and the orbital period $P$ as free parameters in the GP regression. This set of parameters comprises the {\it system parameters}, and were included in the calculation of the eclipse model. 

It should be noted that we used the nearest preceding transit center time ($T_C$), period, \esinw, and \ecosw to predict the eclipse center time, $T_S$. $T_C$ was used as a free parameter in the regression to account for inaccuracies in the linear ephemeris. The eclipse depth and transit center time are the {\it eclipse parameters}\footnote{Technically, the system and eclipse parameters are also hyperparameters of the GP as defined by \cite{gibson12}. However, we refer to the system and eclipse parameters separately for clarity.}.

In addition to the likelihood due to specific hyperparameters and model parameters, we included additional likelihoods based on prior previous measurements of the system and logical constraints on hyperparameters. In our GP regression, we therefore maximized the combined prior and model log-likelihood function

\begin{equation}
	\ln(P) = \ln(L_{\rm model})+\ln(L_{\rm prior})
	\label{eq:lnp}
\end{equation}

\noindent where $P$ is the complete set of parameters used in the GP regression, $L_{\rm model}$ is the likelihood from the systematic and light curve model, and $L_{\rm prior}$ is the likelihood of the prior information. The prior probability distributions were chosen to be uniform, normal, or have no restrictions. 

\subsection{Comparing GP to Parametric Regression}
\label{ref:compare}

GP regression is a relatively new technique as applied to exoplanet spectra, and thermal emission spectra in particular \citep{hd209gp,montet16}.  Exo-atmosphere spectrophotometry is typically plagued with instrumental systematics and noise that can only partially be attributed to known physical sources, such as the hook seen in {\it HST}/WFC3 scans being attributed to charge trapping in the detectors \citep{wfc3ir,corot2b}. The physical origins of other observed systematics are unknown. Even in cases where some observed systematic effect can be attributed to a physical source, we lack understanding as to {why} the systematic can be modeled by a particular parametric form. 

Additionally, when modeling transit or eclipse light curves with a parametric approach, great care needs to be taken when attempting to combine data from multiple visits or multiple scan directions. The instrumental noise is likely different between visits and directions, and additional parameters like a multiplicative flux offset need to be included. When using the parametric approach, most studies tend to treat multiple visits and different scan directions separately, which can reduce the S/N of the light curves. The effects of common-mode (white-light) noise need to be removed from the spectrally resolved light curves before fitting to ensure that uncertainties due to common-mode noise are not being counted twice, and overestimating the uncertainties on eclipse depth. This is sometimes done by first calculating differential light curves for each wavechannel by dividing the spectrally resolved light curve by its corresponding white-light curve (preserving visit and scan direction). Each new source of noise accounted for in a parameterized regression will add a handful of new free parameters, which can quickly become computationally challenging with many visits, orbits, and scan directions.

Most of these details become much less crucial, or altogether irrelevant, when GP regression is applied to the problem. The individual sources of noise or systematic effects (e.g. common-mode, read noise, photon noise, background subtraction, charge trapping, etc.) do not need to be explicitly considered in a GP regression. Rather, GP regression deals with the cumulative effect of all potential sources of noise on the measured light curve, and simply requires that the chosen covariance kernel is flexible enough to account for any behavior not determined by the eclipse model \citep{gibson12,gibson12b}. This prevents double-counting of common-mode noise, over- or underparameterization of noise, and unknown sources of noise.

Combining different scan directions becomes trivial with GP regression, as any flux offset between the directions (which would require an additional free parameter in a parametric fit) is implicitly accounted for in the predictive mean (noise+eclipse model) of the GP regression. We verified this by first fitting the forward- and backward-scanned white-light curves separately, then combined, and compared the solutions. We found that the forward- and backward- scan solutions produced nearly identical hyperparameter solutions, but that the uncertainties on the individual fits were larger than the combined fit. When combined, any minor differences in the noise solutions of the two directions were accounted for by the covariance kernel, and the resulting uncertainties were smaller. We still chose to fit the two {visits} separately to demonstrate the repeatability of our measurements; if that had not been a concern, we might have combined data from two visits with similarly improved results.

We likewise tested whether fitting differential light curves instead of the spectrally resolved light curves improved the precision of the GP regression. We found that when using the same covariance kernel for the differential and non-differential fits, the resulting spectra differed by an average of only 6 ppm and had identical shapes and slopes. However, the eclipse depth uncertainties were 3\% larger for the differential fits. The residuals to the differential fits also still exhibited more non-normal behavior than the non-differential fits. As the differential light curves lack the periodicity observed in the white-light curves by design, it is logical that the quasi-periodic kernel is may no longer be appropriate. It is possible that regression of differential light curves using different covariance kernels might yield a precision that surpasses the non-differential fits, but they may run the risk of overfitting the data and attributing all variation to noise and none to the actual eclipse. While differential light curves present a distinct advantage when using parametric regression techniques, that advantage is not needed with GP regression provided an appropriate kernel is used, and so we fit the non-differential spectrally resolved light curves. 

\cite{ingalls16} tested the repeatability and accuracy of various exoplanet eclipse fitting techniques, including GP regression, using real and simulated {\it Spitzer} light curves of XO-3b. When compared to other fitting techniques, such as nearest neighbor kernel regression, pixel level decorrelation, and independent component analysis, GP regression produced eclipse depths that were consistent with other techniques but had larger uncertainties on those depths (see Fig. 8 and Tables 3 and 4 of \citealt{ingalls16}). However, other techniques that produced {inconsistent} eclipse depths had very small uncertainties. This highlights a key benefit of GP regression: a good fit produces realistic solutions and uncertainties, while a poor fit produces unrealistic parameters and unrealistic uncertainties. In short, GP produces either obvious correct answers or obvious incorrect answers. With other, less flexible methods, an incorrect answer can still have small uncertainties, which could lead to false confidence in a poor result.

We note that GP can become computationally expensive for larger datasets, as each attempt at solving Eq. \ref{eq:gplike} requires inversion of an $n\times n$ matrix to obtain the likelihood value. Calculating the predictive mean requires inverting a $n\times n_{\rm test}$ matrix, with $n_{test}$ as the number of times a measurement is to be predicted, typically $\sim10\times n$. For datasets with hundreds or thousands of measurements, parallel computing is necessary, or application of sparse GP methods (e.g. \citealt{sparsegp1,sparsegp2}).

While the GP regression should accurately account for all undesired detector behavior, we also fit our observations using a traditional parametric approach to verify that our results were not dependent on methodology. We parameterized the data with an exponential+linear trend model similar to \cite{knutson14b}, given by

\begin{equation}
	F(t) = c_1 \left(1+c_2*t + \sum_{i=2}^5 c_{3,i}*e^{\sfrac{-p_i}{c_{4,i}}}\right)*F_{\rm LC}(t)
\label{eq:param}
\end{equation}

\noindent where $t$ is the time from the start of observations, $p_i$ is the time from the start of orbit $i$, $c_1$ and $c_2$ characterize the behavior across the duration of the observation, $c_{3,i}$ and $c_{4,i}$ characterize the behavior of each orbit $i$, and $F_{\rm LC}(t)$ is the eclipse light curve given by \cite{mandel}. Each scan direction and visit was considered separately. We used bootstrap resampling to obtain uncertainties in the parametric fit in Eq. \ref{eq:param}, and a Markov chain Monte Carlo (MCMC) to obtain the uncertainties in the GP regression in Eq. \ref{eq:lnp}.

\begin{figure}[t]
	\begin{center}
	\includegraphics[width=0.45\textwidth]{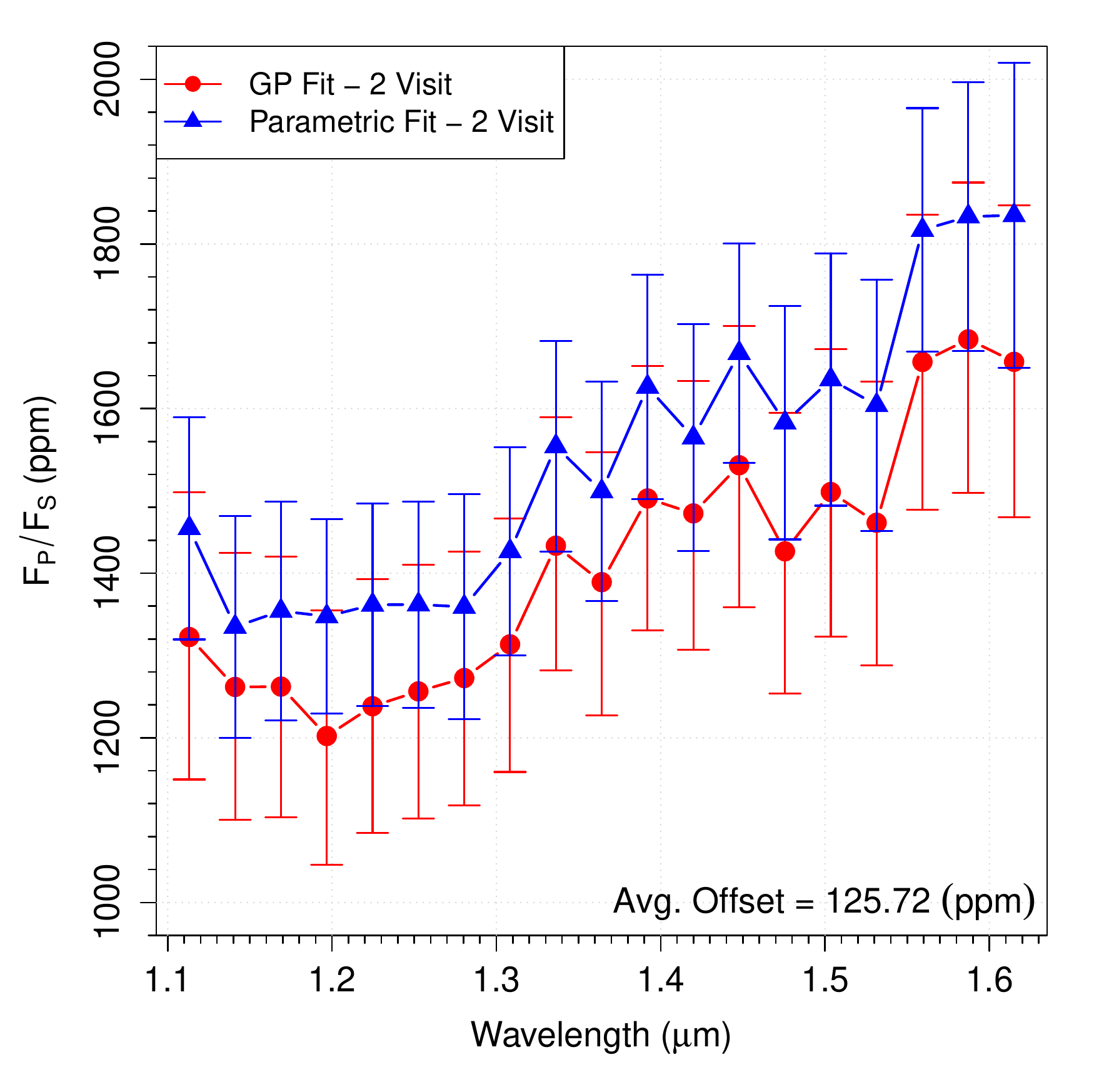}
	\caption{Visit-averaged spectra generated using parametric fitting (blue triangles and line) and GP regression (red points and line). Both spectra have been decontaminated for flux from the companion star. The spectra produced by each technique are both featureless and have the same slope, and the parametric regression indicates an overall hotter planet.}
	\label{fig:compare}
	\end{center}
\end{figure}

Fig. \ref{fig:compare} compares the thermal emission spectrum resulting from each fitting technique for the scan-direction-combined visit-averaged light curves. We first combined the scan directions in each visit, then combined each visit to generate a visit-averaged spectrum via parametric regression. These spectra were also corrected for flux from the companion star.  We found that our parametric regression and our GP regression produced spectrum behavior and eclipse depths consistent within $1\sigma$, but that the parametric fit produced larger eclipse depths than the GP regression at an average offset of $+125.72$ ppm. The RMS of the residuals to the GP regression were only slightly larger than the parametric regression (GP: 526 ppm; parametric: 521 ppm), so we know that the offset cannot be attributed to the quality of each regression. 

Examination of the normality of the residuals to the (scan and visit-separated) parametric fits reveals five wavechannels that fail the Anderson-Darling normality test at the highest significance level; residuals to only one wavechannel fail the normality test in the case of GP regression. We therefore interpret the eclipse depth offset between techniques to mean that the data contain non-Gaussian noise that remains unaccounted for by the parametric model, so that additional variation was instead attributed instead to the eclipse. The GP regression produced slightly larger uncertainties in the eclipse parameters (GP: 175 ppm; parametric: 139 ppm), which was in line with our expectations for GP regression. 

We verified that the GP did not overestimate the uncertainties by comparing the uncertainties on the eclipse depth generated with GP for each spectral wavechannel to what would be expected if the data had perfectly white noise. For perfectly white noise, we would expect uncertainties equal to 

\begin{equation}
\label{eq:photon}
\sigma_{white}={\rm RMS(residuals)}/\sqrt{n/2}
\end{equation}

\noindent where RMS(residuals) is the root-mean-square of the residuals to the GP regression and $n$ is the number of data points for each light curve ($n=88$ for each wavechannel). If we had observations with perfectly white noise added, we would expect to obtain eclipse depth uncertainties of $\sim70$ ppm. We find that, on average, the eclipse depth uncertainties are $2.5\times$ greater than the expected white-noise uncertainty (without accounting for contamination from the companion star). It is clear, however, that the data show time-correlated noise, and accounting for the presence of this correlated noise inflates the GP depth uncertainties up by a factor of $2.5$. 

While the uncertainties produced by GP regression are typically larger than those produced by parametric regression, these uncertainties will account for both white and time-correlated noise in the data, and will therefore be more accurate. We report the results of our GP regression for the remainder of this manuscript.

\subsection{Regression Procedure}
\label{sec:regression}

Here we describe the machinery of our GP regression procedure, the application of which is described in Sec. \ref{sec:whitespec}.

We extracted the aperture-optimized eclipse light curves from the forward and backward scans of each visit to use in our regression. We found that modeling the forward-scan direction and the backward-scan direction separately yielded worse fits, therefore we combined the forward and backward scans for the entirety of the regression. We did not combine light curves from the two visits because potentially different detector systematics visit-to-visit might not have been corrected by the GP, also to investigate the repeatability of our results. Before combining scan directions, we trimmed the light curves to remove the first orbit of each visit, which captured the greatest detector systematics, and trimmed the first point in each remaining orbit for the same reason. Each light curve contained 88 flux measurements after trimming and combining scan directions.

We defined priors on all hyperparameters and system parameters, the values and widths of which were the same for each visit for the white-light eclipses (except for $T_C$). For the spectral data we used priors based on the results of each visit's white-light results. We set the prior value of the characteristic timescale of the periodicity, $\theta$, to match the timescale of an {\it HST} orbit ($\theta=0.06628\pm0.00003$ day) for all light curves, knowing {\it a priori} that this is the timescale on which the systematic trend repeats. We determined the $\theta$ prior from observations of the {\it HST}'s orbital elements, provided by archival two-line element sets for {\it HST} for the dates of our two visits obtained by emailing the United States Joint Functional Component Command for Space.

{While this prior is very restrictive, in initial tests where we allowed $\theta$ more freedom to vary, the posterior on $\theta$ still converged on the orbital periodicity. However, when coupled with similarly unrestricted priors for $A,\,L,$ and $\Omega$, the GP sometimes attempted to fit the noise rather than the eclipse, reported erroneously small eclipse depths, and yielded very poor fits on other parameters. Since a prior value and uncertainty on $\theta$ is physically motivated by our system (but not so for other hyperparameters), we adopted the listed value to gain higher confidence in the GP regression and allow freer exploration of the other hyperparameters. We are confident that restricting $\theta$ to the \textit{HST} orbital timescale does not aversely affect the regression.}

Given these prior values, we used an {\tt amoeba} maximization to find an initial best fit to the eclipse light curve, the results of which were then fed as starting values into an MCMC sampling of Eq. \ref{eq:lnp}. We used the differential evolution Monte Carlo algorithm from the {\tt EXOFAST} package \citep{exofast} to explore the parameter space around the {\tt amoeba} solution. The median and $1\sigma$ values from the MCMC chains were used to calculate the predictive mean (eclipse model and systematics) of the likelihood function defined in Eq. \ref{eq:gplike}. 

\subsection{White-light and Spectral Regression}
\label{sec:whitespec}

\begin{table}[t]
\renewcommand{\arraystretch}{1.3}
\begin{center}
\caption{Prior values, widths, and shapes for white-light regression}
\label{tab:priors}
\begin{tabular}{cccc}
\hline
Parameter			&	Value$\pm$Width			&	Distribution Type\\  \hline\hline 
$A$					&  $0.1\pm0.1$ 				&	Unrestricted\\
L (day)				&	$\tau_{14,s}/2\le L\le20$&	Uniform\\
$\theta$ (day)		&	$0.06628\pm0.00003$	&	Normal\\
$\Omega$ (day)	&	$\tau_s\le\Omega\le20$	&	Uniform\\ \hline
${\rp}/{\rs}$		&	$0.1158\pm0.0006$		& 	Normal		\\
${\aplan}/{\rs}$	&	$2.9398\pm0.03$			&	Normal \\ 
$\cos i$				&	$0.032\pm0.017$		&	Normal	\\
\ecosw				&	$0\pm0.01$					&	Normal	\\
\esinw				&	$0\pm0.01$					&	Normal	\\
$\rs$ (\rsun)		&	$1.419\pm0.055$			&	Normal	\\
Period (day)			&	$0.9255\pm0.00002$	&	Normal		\\ \hline
Eclipse Depth (ppm)&$0.001\pm0.001$			&	Unrestricted		\\
$T_{C,V1}$ (JD)	&	$2457188.923\pm0.001$&	Normal\\
$T_{C,V2}$ (JD)	&	$2457190.774\pm0.001$ &	Normal\\
\hline
\end{tabular}
\end{center}
\vspace{-0.3cm}
\tablecomments{All priors are the same for Visit 1 and Visit 2 except where explicitly indicated with ``V1'' and ``V2.'' For Distribution Type ``Unrestricted,'' no prior limits were placed on this parameter, and the listed values were used as starting points and scale lengths for the {\tt amoeba} maximization.}
\end{table}
\renewcommand{\arraystretch}{1}

\begin{table}[t]
\renewcommand{\arraystretch}{1.3}
\begin{center}
\caption{White-light Solutions from GP Regression}
\label{tab:sysparams}
\begin{tabular}{ccc}
\hline
Parameter			&	Visit 1							&	Visit 2 \\ \hline\hline
A						&	$0.141\pm0.19$			&	$0.049\pm0.045$\\
L (day)				&	$9.8\pm3.4$				&	$10.3\pm3.4$\\
$\theta$ (day)		&	$0.06628\pm0.00004$	&	$0.06628\pm0.00004$\\
$\Omega$ (day)	&	$9.7\pm3.4$				&	$9.6\pm3.4$\\ \hline
${\rp}/{\rs}$		&	$0.1158\pm0.0008$		&	$0.1158\pm0.0009$		\\
${\aplan}/{\rs}$	&	$2.86\pm0.048$			&	$3.006\pm0.058$			\\
$\cos i$				&	$0.025\pm0.021$			&	$0.035\pm0.026$	\\
\ecosw				&	$-0.003\pm0.014$		&	$-0.039\pm0.025$		\\
\esinw				&	$0.000\pm0.014$			&	$-0.001\pm0.018$		\\
$\rs$	(\rsun)		&	$1.436\pm0.056$			&	$1.436\pm0.057$		\\
Period (day)			&	$0.92555\pm0.00003$	&	$0.92555\pm{0.00003}$	\\ \hline
Eclipse Depth (ppm)&$1246\pm63$				&	$1196\pm49$			\\
$T_S$ (JD)			&	$2457189.38\pm0.0008$&	$2457191.23\pm0.0012$	\\ \hline
$T_C+t_*$ (JD)	&	$2457188.92\pm0.0008$&	$2457190.99\pm0.0008$ \\ 
$T_{14,S}$ (day)&	$0.1177\pm0.0021$		&	$0.1113\pm0.0012$\\
$\tau_S$ (day) 	&	$0.0128\pm0.00027$	&	$0.0121\pm0.00033$  \\ \hline
\end{tabular}
\end{center}
\vspace{-0.3cm}
\tablecomments{The eclipse depth and uncertainties have not been decontaminated for flux from the companion star.}
\end{table}
\renewcommand{\arraystretch}{1}


We first fit secondary eclipse light curve models to the blended white-light data for each visit as described in Sec. \ref{sec:regression}. In the initial fits, we used as priors the \cite{southworth} values for ${\rp}/{\rs}$, $\cos i$, orbital period, and ${\aplan}/{\rs}$ and assumed a near-circular orbit with small argument of periastron. The prior values, widths, and shapes for all hyperparameters, system parameters, and eclipse parameters are listed in Table \ref{tab:priors}. $\rs$ together with ${\aplan}/{\rs}$ was used to calculate and correct for the Roemer delay (i.e. the light travel time) across the orbit to accurately determine the secondary eclipse time. The lower limits on $L$ and $\Omega$ were chosen to be half of the transit duration ($T_{14,s}/2=0.054$ day) and the ingress/egress duration ($\tau_s$=0.01 day), respectively. This ensured that the GP regression modeled the systematics across the orbit instead of fitting the scatter between the points. The upper limits on $L$ and $\Omega$ ensured that these hyperparameters did not become so large that changes to these hyperparameters dominated the calculation of $ln(P)$.

\cite{southworth} found that \planet\, is slightly aspherical as a result of its close-in orbit. It has a Roche-lobe filling factor of 0.58 and an equatorial radius 2.2\% larger than the polar radius.  We tested whether assuming a spherical or aspherical prior value for \rp\, affected the measured eclipse depth while using a spherical eclipse model, i.e. how robust the GP regression is to small changes in \rp\ prior without changing the eclipse model. We fit the light curve model to the white-light data of each visit via GP regression and MCMC sampling using both the best-fit spherical and aspherical radius values from \cite{southworth} as priors for the regression. In comparing the two solutions, we found that the residuals {within} each fit are $374.1$ times greater than the residuals {between} the the two solutions. From this we conclude that any impact on our regression caused by the planetary asphericity is accounted for in the MCMC sampling. We set the prior value of the planetary radius to the spherical value,$\rp=1.554\pm0.044$, for simplicity.

We fit the blended white-light data for each visit at the optimum aperture via GP regression with MCMC sampling using the priors listed in Table \ref{tab:priors}. The eclipse, system, and hyperparameter solutions from GP regression of the white-light eclipses are listed in Table \ref{tab:sysparams}. We used the white-light solutions to calculate the corrected transit center time, $T_C+t_*$, eclipse duration, $T_{14,S}$, and the eclipse ingress/egress duration, $\tau_S$, for completeness. 

We find that the white-light solutions for Visit 1 and Visit 2 are consistent with each other within $1\sigma$, with similar uncertainties on parameters for each visit. We find that except for the characteristic timescale of periodicity in the covariance kernel ($\theta$), which was known very precisely {a priori}, each of the hyperparameters has relatively large uncertainties. This is because once the timescale for $L$ or $\Omega$ greatly exceed the duration of our observations, the covariance matrices generated using these large hyperparameter values are degenerate within the timescale of our data. For example, the covariance between our first and last observations, which are separated by $\sim0.29$ day, is effectively the same for covariance timescales of $L=5$ and $L=15$. However, as long as $L$ and $\Omega$ remain above their lower limit values, we find that the uncertainties on these hyperparameters to not significantly impact the precision of the eclipse depth\footnote{For regression attempts where $L$ or $\Omega$ was allowed to reach values below the lower limit listed in Table \ref{tab:priors}, the GP model effectively attributed all inter-orbit variations to noise and not to the eclipse itself, resulting in anomalously small eclipse depths.}. {As expected, the posterior widths for nearly all fitted parameters, including $\theta$, are slightly wider than the prior widths, as the GP and MCMC compound the parameter uncertainties regardless of the prior distribution shape (within allowed boundaries).}

\begin{figure*}[t]
	\begin{center}
	\includegraphics[width=\textwidth]{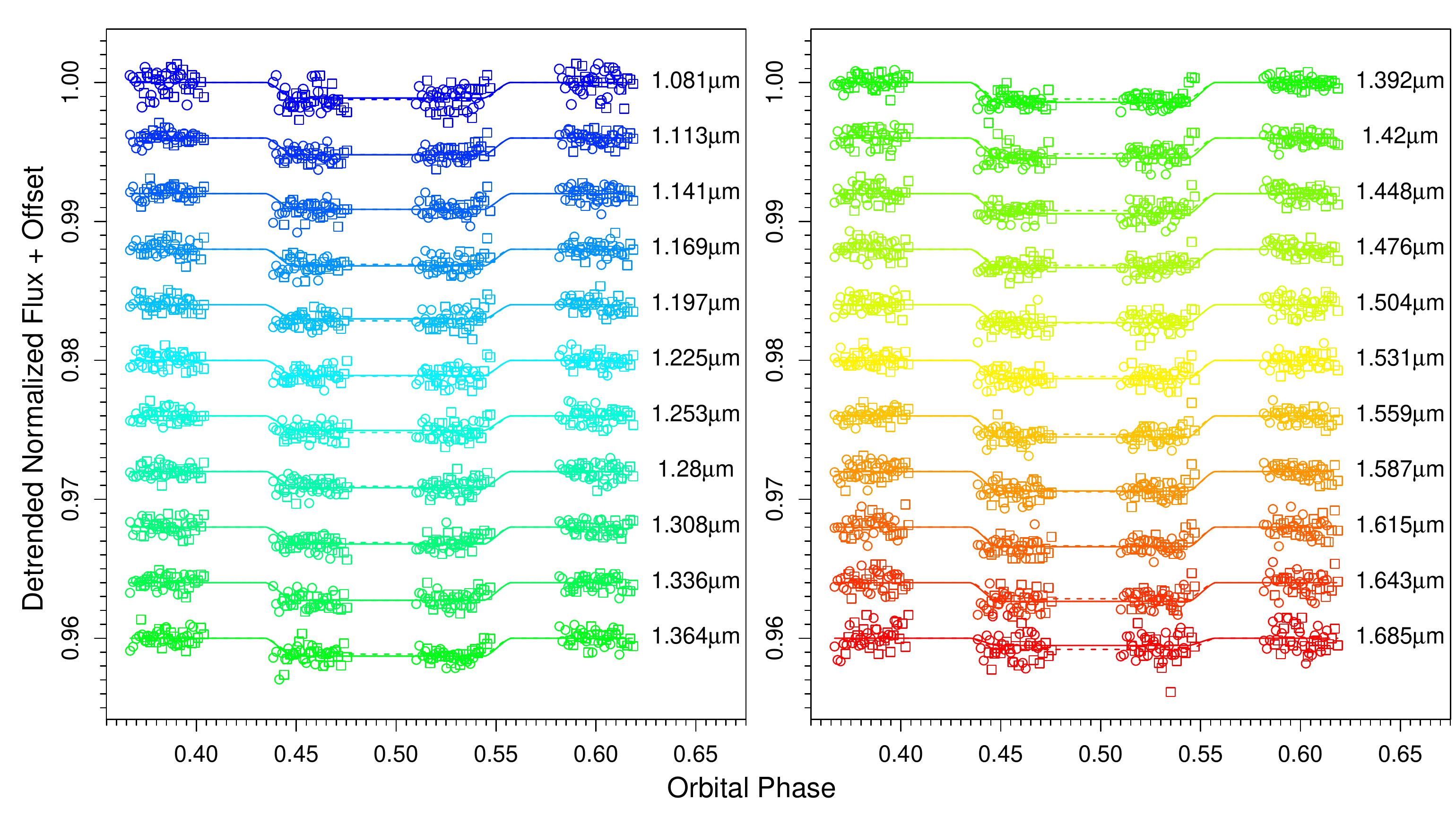}
	\caption{Secondary eclipse light curves for spectral wavechannel data detrended for detector systematics via GP regression, before correction for flux contamination. Open circles and solid line indicate Visit 1 data and model, respectively, and open squares and dashed lines indicate Visit 2 data and model, respectively. A vertical offset is added for clarity. Wavechannels are color coded and labeled by the visit-averaged wavelength, $\overline{\lambda}$.}
		\label{fig:spectral}
	\end{center}
\end{figure*}

We used the white-light solution at the optimum aperture as starting values and prior probabilities for fitting the binned spectral data via GP regression. We used the same priors on $L$ and $\omega$ as used in the white-light fit. As the binned spectral data do not have much leverage on the system parameters, we held all system parameters (${\rp}/{\rs}$, ${\aplan}/{\rs}$, $\cos i$, \ecosw, \esinw, and $\rs$) fixed at their white-light values for the regression (Table \ref{tab:sysparams}), and only fit the four hyperparameters ($A,\,L,\,\theta$, and $\Omega$) and the two eclipse parameters (eclipse depth and eclipse center time) for each of the 22 spectral wavechannels as described in Sec. \ref{sec:regression}. 
We also combined forward- and backward-scan directions in each wavechannel to improve the S/N of the data. 


Fig. \ref{fig:spectral} shows the
 eclipse light curves in each wavechannel with the detector systematics removed and the best-fit eclipse models overplotted. As seen in Fig. \ref{fig:spectral}, the GP regression was able to capture the correlated noise in the eclipse light curves, which allowed for more precise light-curve regression.
 



\subsection{Flux Decontamination}
\label{sec:contam}

Before constructing the final thermal emission spectrum, we corrected each wavechannel for contamination from the companion star as described in Sec. \ref{sec:second}. The eclipse depths obtained through the GP regression do not represent the true planet/star flux ratio until after flux decontamination. We extracted flux contamination ratios at the central wavelengths of each wavechannel on each visit from the SED models for the primary and companion components, accounting for slight differences in wavelength solutions for each visit. The contamination ranged between $\sim9\%$ for the shortest wavelengths and $\sim17\%$ for the longest wavelengths. The decontaminated eclipse depth yields the true planet/star flux ratio, ${F_P}/{F_S}$, which is given by

\begin{equation}
\frac{F_P}{F_S}(\lambda) = {\rm d_{\lambda}}\,\times\left[1-\frac{F_B}{F_A}(\lambda)\right]^{-1}
\label{eq:contam}
\end{equation}

\noindent where ${\rm d_{\lambda}}$ is the eclipse depth before flux decontamination
 and ${F_B}/{F_A}(\lambda)$ is the fractional contribution of flux from the companion star at wavelength $\lambda$. The flux contamination ratios for each visit in each wavechannel are listed in Table \ref{tab:contam}. The contamination ratios we calculate are consistent with the values reported by \cite{newsouth}, who calculate the contamination from the companion star for Bessel $RI$ and $griz$ passbands. We note that with this method were are simply scaling the eclipse depths and uncertainties by the contamination ratio, and do not incorporate the added (minor) uncertainty of the flux contamination ratio itself.


\begin{table}[t]
\begin{center}
\caption{Flux contamination ratio due to the companion star for Visit 1 and Visit 2}
\label{tab:contam}
\begin{tabular}{cc|cc}
\hline
\multicolumn{2}{c|}{Visit 1}	&	\multicolumn{2}{c}{Visit 2} \\ \hline
$\lambda_1\,(\micron)$	&	${F_B/F_A (\lambda_1)}$	&	$\lambda_2\,(\micron)$	&	${F_B/F_A (\lambda_2)}$\\  \hline\hline
1.0783	&	0.0922	&	1.0829	&	0.0923	\\
1.1108	&	0.0952	&	1.1155	&	0.0955	\\
1.1387	&	0.0965	&	1.1433	&	0.0968	\\
1.1666	&	0.0997	&	1.1712	&	0.1004	\\
1.1945	&	0.1016	&	1.1991	&	0.1015	\\
1.2223	&	0.1050	&	1.2270	&	0.1055	\\
1.2502	&	0.1074	&	1.2549	&	0.1080	\\
1.2781	&	0.1118	&	1.2828	&	0.1163	\\
1.3060	&	0.1132	&	1.3107	&	0.1126	\\
1.3339	&	0.1158	&	1.3385	&	0.1167	\\
1.3618	&	0.1186	&	1.3664	&	0.1199	\\
1.3897	&	0.1228	&	1.3943	&	0.1242	\\
1.4175	&	0.1273	&	1.4222	&	0.1279	\\
1.4454	&	0.1315	&	1.4501	&	0.1321	\\
1.4733	&	0.1352	&	1.4779	&	0.1364	\\
1.5012	&	0.1360	&	1.5058	&	0.1376	\\
1.5291	&	0.1454	&	1.5337	&	0.1463	\\
1.5570	&	0.1508	&	1.5616	&	0.1512	\\
1.5848	&	0.1547	&	1.5895	&	0.1559	\\
1.6127	&	0.1600	&	1.6174	&	0.1602	\\
1.6406	&	0.1639	&	1.6453	&	0.1647	\\
1.6824	&	0.1660	&	1.6871	&	0.1662	\\ \hline

\end{tabular}
\end{center}
\end{table}

\subsection{Normality of Residuals}
\label{sec:rnorm}

\begin{figure*}[t]
	\begin{center}
	\includegraphics[width=0.95\textwidth]{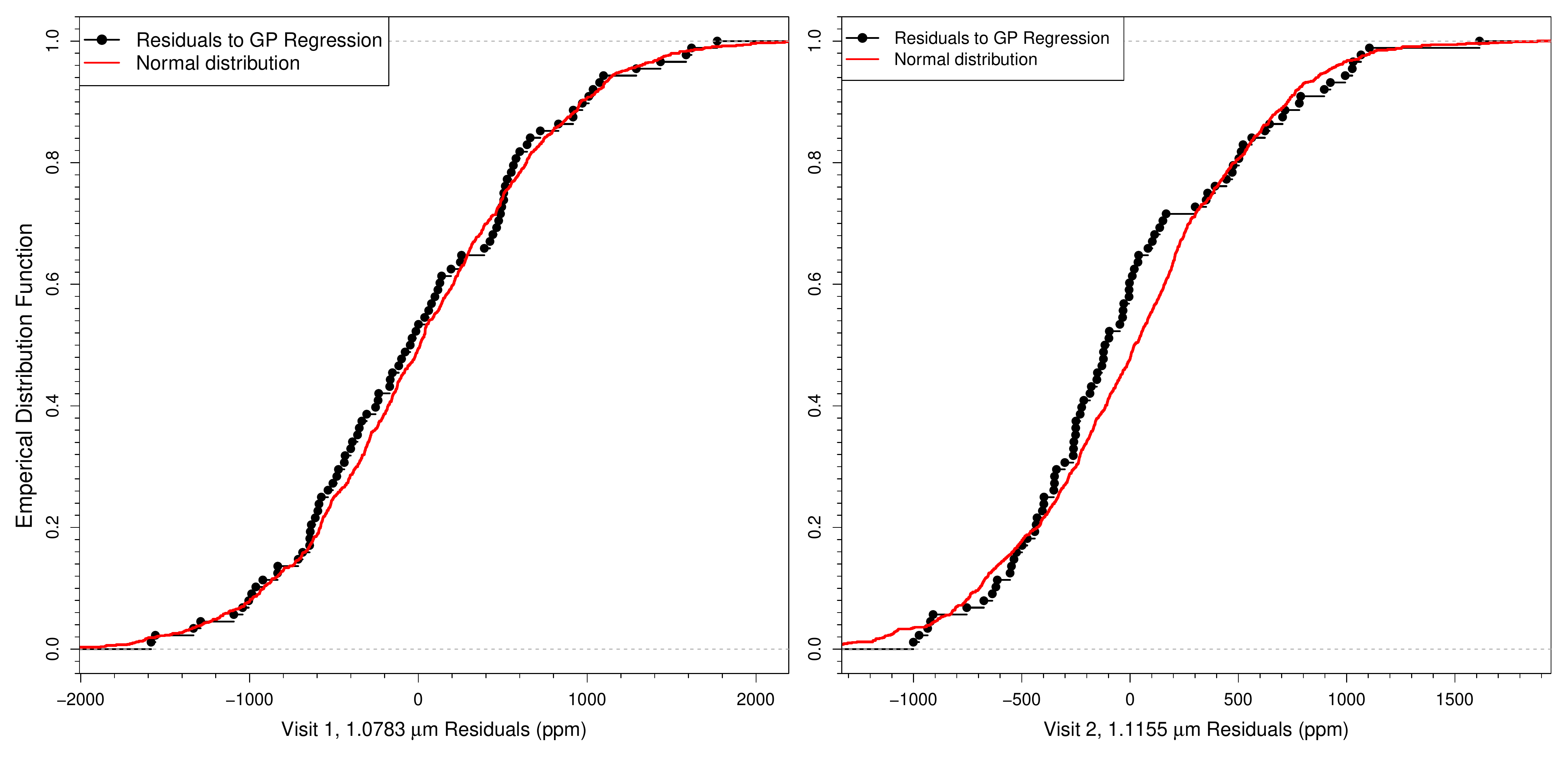}
	\caption{Left: empirical distribution function of residuals to Gaussian process regression of Visit 1, $\lambda=1.0783\,\micron$ light-curve in ppm (black connected dots), which passes the A-D test for normality at 10\% significance, compared to that of a normal distribution (red solid line). Right: same as left for Visit 2, $\lambda=1.1155\,\micron$, which does not pass the A-D test at 1\% significance.}
	\label{fig:edf}
	\end{center}
\end{figure*}

For a GP regression to be deemed successful, the residuals to the model should not contain any non-Gaussian behavior. We therefore tested the residuals of the spectral GP regression for normality using the Anderson-Darling test (A-D test; \citealt{adtest,astrostats,nortest}). The A-D test states that if the A-D statistic, $A^2$, is above a critical value, then the hypothesis that the data are drawn from a normal distribution is rejected at a specified significance level. A significance level of $\alpha=0.01$ (1\% significance) corresponds to the probability of observing the tested phenomenon by chance. The critical values depend on the number of points in the sample and the desired significance level of the result. 

We adjusted the A-D statistic for the unknown mean and variance of the prior distribution (i.e. a Case 3 A-D test) using

\begin{equation}
	A^{*2} = A^2 \left(1+\frac{0.75}{n}-\frac{2.25}{n^2}\right),
\end{equation}

\noindent where $A^2$ is the unadjusted A-D statistic and $n$ is the number of points in the sample. For our spectral GP regression, $n=88$ for each light curve after clipping the first orbit, trimming the first point in each remaining orbit, and combining the forward and backward scans. 

We computed the adjusted A-D statistic for the white-light and each wavechannel in Visit 1 and Visit 2 for 10\% ($A_{\rm crit} = 0.6287$), 5\% ($A_{\rm crit} = 0.7468$), and 1\% ($A_{\rm crit} = 1.0379$) significance levels. The A-D test indicates that we cannot reject normality for white-light residuals for either visit at the $1\%$ significance level. The A-D test further indicates that we cannot reject normality at the $1\%$ significance level for any wavechannel except $\lambda=1.5058\,\micron$ in Visit 2 ($A^{*2}=1.069$), and the normality of only a few wavechannels is rejected at the $5\%$ or $10\%$ level in either visit. When comparing the empirical distribution function (EDF) of the residuals for Visit 2 $\lambda=1.5058\,\micron$ to the EDF of a normal distribution (Fig. \ref{fig:edf}, right), we clearly see the deviation from normality when compared to the EDF of a wavechannel that passes the A-D test at high significance level (Fig. \ref{fig:edf}, left). As Visit 2 $\lambda=1.5058\,\micron$ is only $0.4401$ higher than the 10\% critical value, we are confident that the effects of non-normality on the residuals of that wavechannel are minimal, and are accounted for in the uncertainties generated from MCMC.

\subsection{Thermal Emission Spectrum of WASP-103b}
\label{sec:tes}

\begin{figure*}[t]
	\begin{center}
	\includegraphics[width=0.95\textwidth]{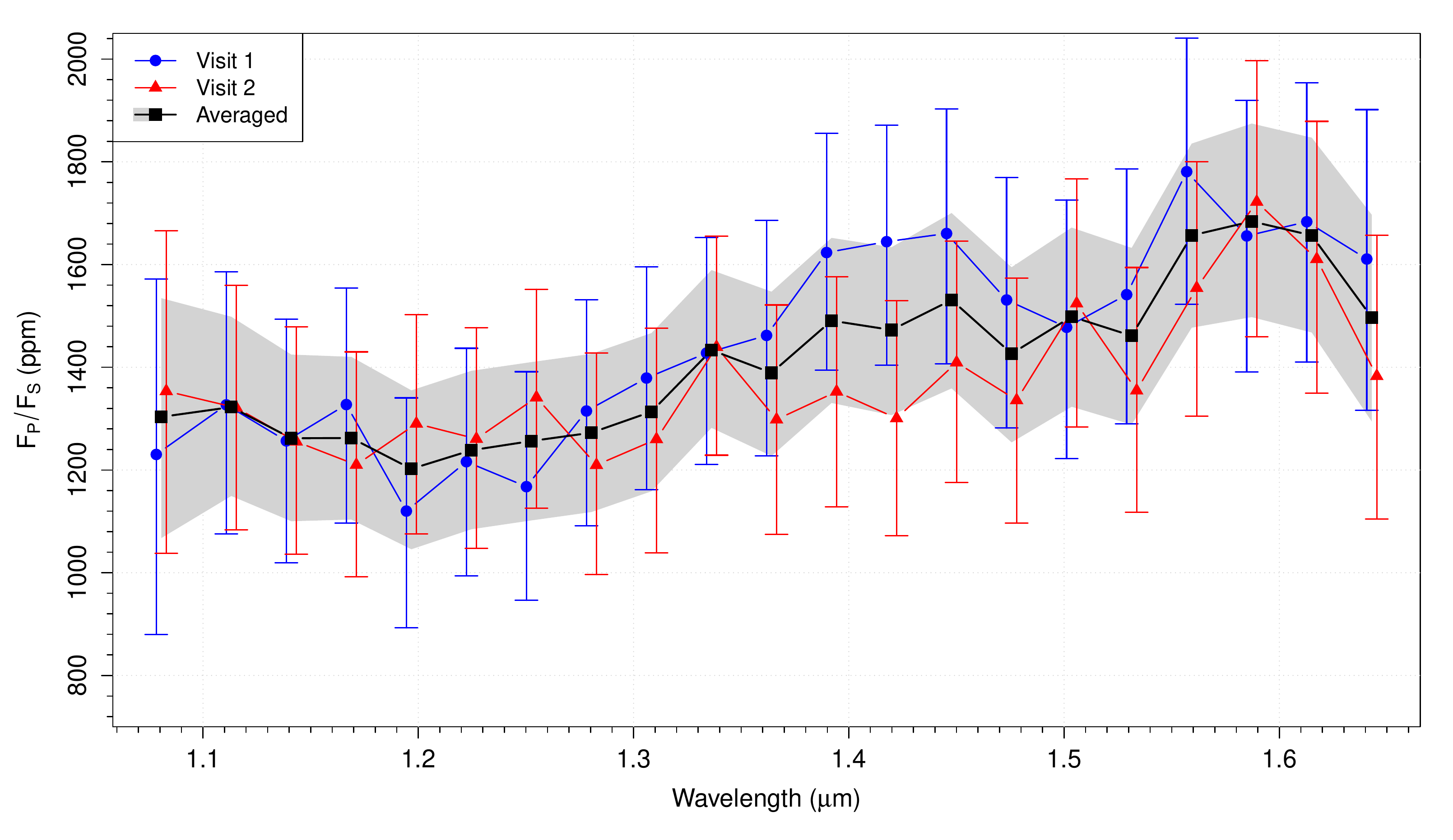}
	\vspace{-0.3cm}
	\caption{Thermal emission spectrum of WASP-103b generated via Gaussian process regression, corrected for flux from the companion star. The blue circles and line indicate spectrum from Visit 1 data, red triangles and line indicate spectrum from Visit 2 data. The black squares are the averaged spectrum over both visits, calculated at the average wavelength of each wavechannel. The gray shaded region is the $1\sigma$ uncertainty of the averaged spectrum found via MCMC.}
	\label{fig:spectrum}
	\end{center}
\end{figure*}

\begin{table}[t]
\renewcommand{\arraystretch}{1.7}
\begin{center}
\caption{Thermal emission spectrum of \planet\, for separate and averaged visits}
\label{tab:tes}
\begin{tabular}{c@{  }c|c@{  }c|c@{  }c}
\hline
\multicolumn{2}{c|}{Visit 1}	&	\multicolumn{2}{c|}{Visit 2}	&	\multicolumn{2}{c}{Averaged} \\
\hline
$\lambda_1$&	$\left({F_P}/{F_S}\right)_{1}$ &$\lambda_2$&	$\left({F_P}/{F_S}\right)_{2}$&$\overline{\lambda}$	&$\overline{{F_P}/{F_S}}$ \vspace{-0.1cm}\\
$\micron$&ppm&$\micron$&ppm&$\micron$&ppm\\ \hline\hline
1.0783&$	1230.3	^{+	341	}_{-	351	}$&	1.0829	&$	1353.6	^{+	312	}_{-	316	}$&	1.0806	&$	1303.4	^{+	231	}_{-	236	}$\\
1.1108&$	1326.8	^{+	259	}_{-	251	}$&	1.1155	&$	1321.3	\pm{		238	}$&	1.1131	&$	1322.5	^{+	176	}_{-	173	}$\\
1.1387&$	1256.7	\pm{		237	}$&	1.1433	&$	1255.2	^{+	223	}_{-	219	}$&	1.1410	&$	1261.8	^{+	163	}_{-	162	}$\\
1.1666&$	1327.4	^{+	227	}_{-	231	}$&	1.1712	&$	1210.0	^{+	220	}_{-	218	}$&	1.1689	&$	1262.3	^{+	158	}_{-	159	}$\\
1.1945&$	1120.1	^{+	220	}_{-	227	}$&	1.1991	&$	1290.3	^{+	212	}_{-	215	}$&	1.1968	&$	1202.2	^{+	153	}_{-	157	}$\\
1.2223&$	1216.1	^{+	221	}_{-	222	}$&	1.2270	&$	1260.7	^{+	217	}_{-	213	}$&	1.2247	&$	1238.5	^{+	155	}_{-	154	}$\\
1.2502&$	1167.6	^{+	224	}_{-	221	}$&	1.2549	&$	1341.3	^{+	211	}_{-	216	}$&	1.2526	&$	1256.5	^{+	154	}_{-	155	}$\\
1.2781&$	1314.9	^{+	217	}_{-	223	}$&	1.2828	&$	1209.7	^{+	218	}_{-	213	}$&	1.2804	&$	1272.6	\pm{		154	}$\\
1.3060&$	1379.0	\pm{		217	}$&	1.3107	&$	1259.9	^{+	216	}_{-	221	}$&	1.3083	&$	1313.6	^{+	153	}_{-	155	}$\\
1.3339&$	1427.7	^{+	225	}_{-	217	}$&	1.3385	&$	1439.5	^{+	216	}_{-	211	}$&	1.3362	&$	1433.4	^{+	156	}_{-	151	}$\\
1.3618&$	1462.0	^{+	224	}_{-	234	}$&	1.3664	&$	1298.3	\pm{		223	}$&	1.3641	&$	1389.1	^{+	158	}_{-	162	}$\\
1.3897&$	1623.4	^{+	232	}_{-	229	}$&	1.3943	&$	1352.7	^{+	223	}_{-	224	}$&	1.3920	&$	1490.8	^{+	161	}_{-	160	}$\\
1.4175&$	1644.6	^{+	227	}_{-	241	}$&	1.4222	&$	1300.7	^{+	229	}_{-	228	}$&	1.4199	&$	1472.6	^{+	161	}_{-	166	}$\\
1.4454&$	1660.5	^{+	242	}_{-	254	}$&	1.4501	&$	1409.3	^{+	236	}_{-	233	}$&	1.4477	&$	1531.1	^{+	169	}_{-	172	}$\\
1.4733&$	1531.1	^{+	239	}_{-	249	}$&	1.4779	&$	1336.0	^{+	237	}_{-	239	}$&	1.4756	&$	1426.6	^{+	168	}_{-	173	}$\\
1.5012&$	1477.8	^{+	248	}_{-	256	}$&	1.5058	&$	1524.2	^{+	242	}_{-	240	}$&	1.5035	&$	1498.6	^{+	173	}_{-	175	}$\\
1.5291&$	1541.5	^{+	245	}_{-	252	}$&	1.5337	&$	1354.7	^{+	240	}_{-	237	}$&	1.5314	&$	1461.3	^{+	171	}_{-	173	}$\\
1.5570&$	1780.8	^{+	260	}_{-	258	}$&	1.5616	&$	1554.4	^{+	246	}_{-	250	}$&	1.5593	&$	1656.7	^{+	179	}_{-	180	}$\\
1.5848&$	1655.7	\pm{		264	}$&	1.5895	&$	1722.1	^{+	275	}_{-	263	}$&	1.5872	&$	1684.1	^{+	191	}_{-	186	}$\\
1.6127&$	1683.4	^{+	270	}_{-	273	}$&	1.6174	&$	1610.6	^{+	268	}_{-	261	}$&	1.6151	&$	1656.8	^{+	190	}_{-	189	}$\\
1.6406&$	1610.8	^{+	291	}_{-	295	}$&	1.6453	&$	1382.8	^{+	274	}_{-	278	}$&	1.6429	&$	1497.0	^{+	200	}_{-	203	}$\\
1.6824&$	622.5	^{+	365	}_{-	321	}$&	1.6871	&$	962.6	^{+	392	}_{-	382	}$&	1.6848	&$	752.6	^{+	268	}_{-	249	}$\\
\hline
\end{tabular}
\end{center}
\vspace{-0.3cm}
\tablecomments{These flux ratios have been corrected for flux from the companion star.}
\end{table}
\renewcommand{\arraystretch}{1.0}

The methods described in Sec. \ref{sec:gp} were applied to the secondary eclipse light curves in each spectral wavechannel observed during each of the two visits with {\it HST} to produce the thermal emission spectrum shown in Fig. \ref{fig:spectrum} and listed in Table \ref{tab:tes}. The longest wavechannel, centered around $1.7\micron$ showed an anomalously low eclipse depth with a very poor fit for both visits. This was likely due to edge effects resulting from the wavechannel binning, and so the final wavechannel was dropped from both Fig. \ref{fig:spectrum} and further analysis. We averaged spectra between the two visits of {\it HST} to calculate the average wavelength, $\overline{\lambda}$, and average planet/star flux ratio, $\overline{{F_P}/{F_S}}$, which were used to retrieve atmospheric models and compare to other exo-atmospheres.

The thermal emission spectrum of \planet\, is featureless across the observed near-IR region down to a sensitivity of 175 ppm, and it exhibits a shallow positive slope toward the red. No significant water absorption is apparent in the $1.4\micron$ water band, nor are any other molecular features present. 

\section{The Atmosphere of WASP-103\lowercase{b}}
\label{sec:atmos}
Here we discuss using the emission spectrum of \planet\ to model the planetary atmosphere, and compare the spectrum of \planet\ to the emission spectra of other exoplanets measured with {\it HST}/WFC3. We also discuss directions for future research that would further help our understanding of the \planet\ atmosphere.

\subsection{Atmospheric Modeling}
\label{sec:atmosmodel}

\begin{figure*}[t]
	\begin{center}
	\includegraphics[width=\textwidth]{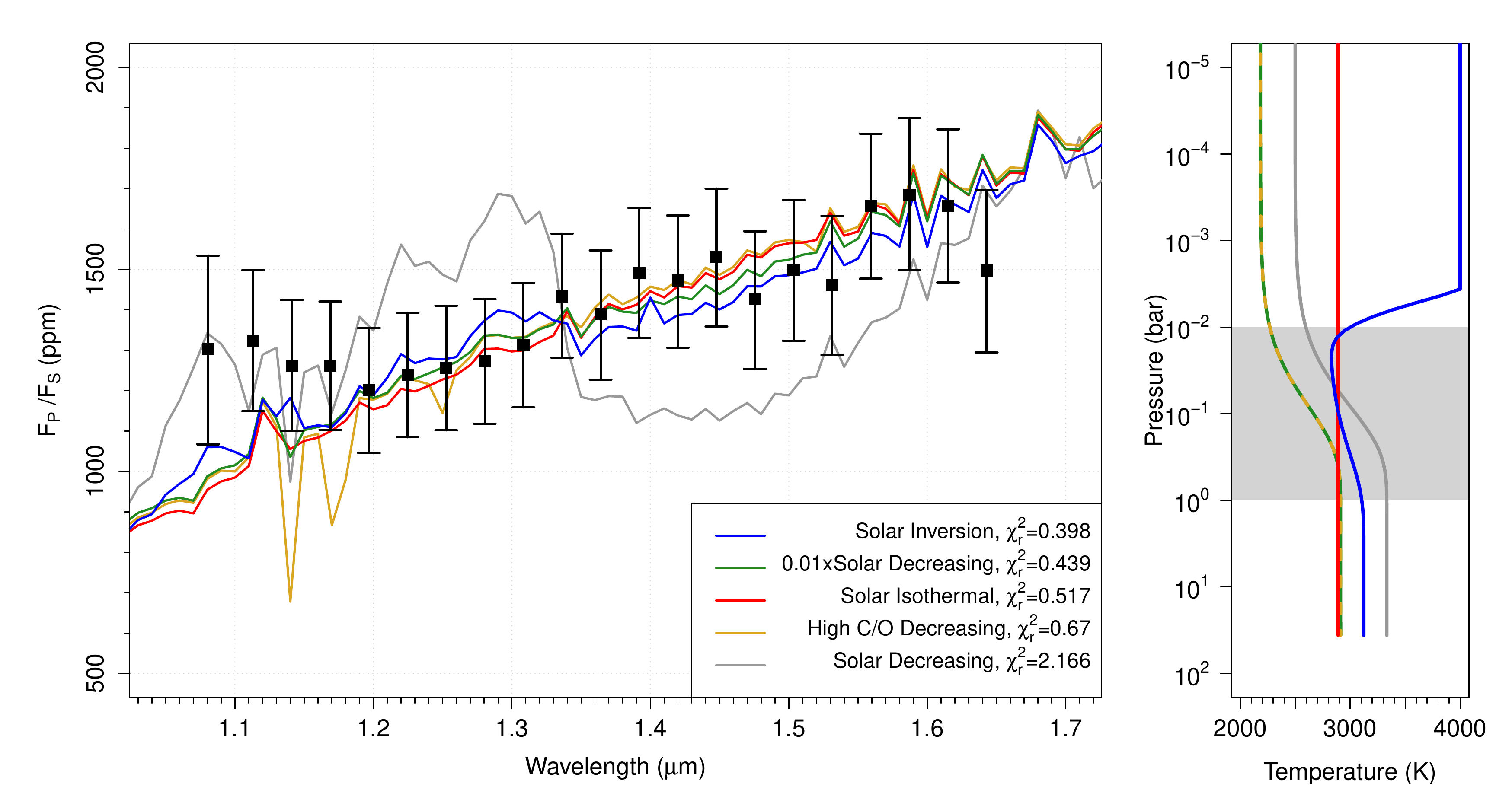}
	\caption{{\it Left:} atmospheric models (binned to low resolution) tested against the visit-averaged thermal emission spectrum of \planet\, in the near-IR (black points). Blue corresponds to a solar metallicity atmosphere with a thermal inversion, yellow corresponds to a decreasing atmosphere with C/O\textgreater1, red corresponds to an isothermal atmosphere at solar metallicity, and green and gray correspond to a decreasing atmosphere at low-metallicity and solar metallicity, respectively.
	Reduced $\chi^2_r$ values are listed for each model. 
	{\it Right:} the vertical pressure-temperature profiles associated with the tested atmospheric models. Model colors are the same as in the left panel, and the gray shaded region indicates the atmospheric pressures probed by our observations. The low-metallicity solar and high C/O profiles are identical, and overplotted with alternating dashed colors.}
	\label{fig:atmos}
	\end{center}
\end{figure*}

We used the thermal emission forward model outlined in \cite{line13a} to retrieve the thermal profile of the \planet\ atmosphere, which uses the \cite{pg14} analytic parameterization of an irradiated non-grey atmosphere. This method uses four parameters to control the ``shape'' of the thermal profile (a visible opacity, two infrared opacities, and the fractional energy split between the two infrared opacities), and one parameter that controls the temperature shift. The relatively featureless nature of the \planet\ spectrum made finding a unique atmospheric model fit to the data difficult. 

{Because of this, we selected a few fiducial atmosphere types that appear frequently in the literature to provide representative solutions. For each atmosphere type, the shape of the temperature profile was held fixed to some standard profile shape, and the fit was reduced to the best temperature shift for that shape. It is possible that better-fitting atmospheric models could be found by iterating shape and shift adjustments (e.g. \citealt{line16}), but {significant} improvement on the model atmosphere fits would require additional data with better leverage on the models.}

We tested our visit-averaged spectrum against five fiducial models: monotonically decreasing atmospheres at solar metallicity, $0.01\times$solar metallicity ($\rm{[Fe/H]}=-2$), and a C/O ratio \textgreater1, an isothermal atmosphere at solar metallicity, and a solar-metallicity atmosphere with a stratospheric thermal inversion (Fig. \ref{fig:atmos}). The monotonically decreasing atmosphere at solar metallicity is rejected via a $\chi^2$ rejection test with 20 degrees of freedom at $\chi^2_r=\chi^2/{\rm d.o.f.}=2.166$.\footnote{The longest wavechannel was left out of the $\chi^2$ tests.} 

The other four scenarios all provide similar, acceptable fits to the spectrum, and therefore we cannot determine which is likely to be correct. Planets are more likely to be enhanced in refractory metals relative to their host star, rather than depleted \citep{ramirez14,thorngren15}. Given the reported near-Solar metallicity of \wasp\,of $\rm{[Fe/H]}=0.06\pm0.13$, it is unlikely that \planet\, is significantly depleted in metals, therefore we disfavor the $0.01\times$solar decreasing model case based on physical, rather than statistical, grounds ($\chi^2/{\rm d.o.f.}=0.439$). The enhanced C/O atmosphere provides an acceptable fit to the spectrum at $\chi^2/{\rm d.o.f.}=0.670$. As no causal link has been established between enhanced C/O and other observable properties of the planetary system, we cannot rule out the enhanced C/O atmospheric model. However, as no other compelling evidence yet exists to suggest that \planet\ has an enhanced C/O ratio, we do not find this model very plausible.

Solar metallicity profiles with an isothermal structure at $T=2890$ K ($\chi^2/{\rm d.o.f.}=0.517$) and a thermal inversion layer near the $10^{-2}$ bar pressure level ($\chi^2/{\rm d.o.f.}=0.398$) provide equally acceptable fits to our spectrum. Across the region of interest, the isothermal and inversion model both show little to no variation in temperature (Fig. \ref{fig:atmos}, right). Given the narrow wavelength range probed and uncertainties of our eclipse depths, we have little power to distinguish between any model that is approximately isothermal in this region. However, the isothermal and thermally inverted models both have a hotter temperature at high altitudes than expected in an monotonically decreasing atmosphere in radiative equilibrium. Since a monotonically decreasing atmosphere should be a good zeroth-order model for an exo-atmosphere, the isothermal and inverted atmospheric models both would require high-altitude absorbers.

Fig. \ref{fig:atmos} (right) highlights that our observations probe an atmospheric pressure at which most models deviate only slightly from the isothermal case, making it extremely difficult to differentiate between models. {While the spectrum may be indistinguishable from isothermal across this wavelength range, it may therefore not manifest from a physically isothermal system, i.e. pseudo-isothermal.} The pseudo-isothermal spectrum indicated by these data could be due to any number of atmospheric phenomena that we cannot detect with our spectrum, including a cloud deck at $P\sim10^{-2}$ bar, high-altitude haze, or a large radiative zone. Alternately, as the right-hand side of Fig. \ref{fig:atmos} suggests, the region probed by our observations may capture the inflection point just below a thermal inversion layer. Our {\it HST} observations have restricted the range of possible thermal profiles, including the altitude of a potential absorber, and indicate a single brightness temperature of ${\rm T_B=2890}$ K across this wavelength range.


If the isothermal or inverted models are correct, then some additional heating is required in the upper atmosphere, which is indicative of some species of higher-altitude molecule that absorbs radiation in the visible and radiates that energy isotropically in the infrared, thus heating the lower layers in the atmosphere. For \planet, this high-altitude absorber is probably TiO \citep{fortney08}, which could be detected through observations probing higher altitudes in the atmosphere (i.e. shorter wavelengths).  Additional eclipse or transit observations at wavelengths shorter than those considered in this study would likely be able to distinguish more clearly between clearly between the enhanced C/O, isothermal, and inverted models and reveal the presence of a high-altitude absorber or other atmospheric phenomena.


\subsection{Comparisons to Other Planets}
\label{sec:discussion}


\begin{figure*}
\begin{center}
\includegraphics[width=0.95\textwidth]{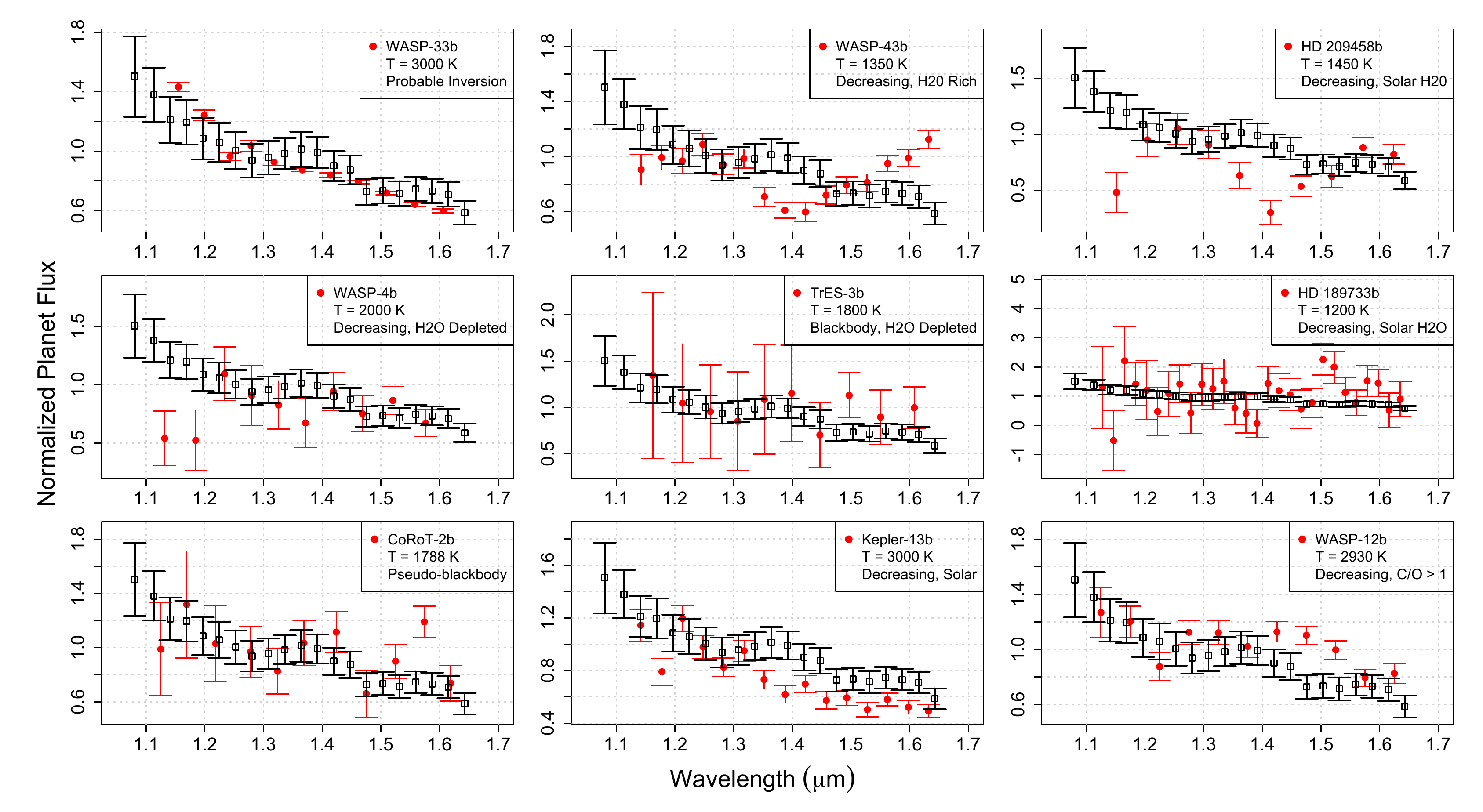}
\caption{Comparisons of normalized planetary spectra taken with {\it HST}/WFC3 G141 across the $1.1-1.7\,\micron$ range. In all panels, the visit-averaged spectrum of \planet\, is shown in open black squares and the comparison planet's spectrum is shown in solid red points. Spectra have been normalized to the average flux value between $1.2-1.3\micron$. Published values of planetary temperature and atmospheric features are included for each comparison planet. References for spectra are found in the text.}
\label{fig:fluxcompare}
\end{center}
\end{figure*}


We compare the planetary spectrum of \planet\ to other exoplanets for which a $1.1-1.7\micron$ emission spectrum has been measured with {\it HST}/WFC3 G141 in Fig. \ref{fig:fluxcompare}. {We calculated the absolute planetary emission spectra by retrieving stellar spectra from the NASA Infrared Telescope Facility Spectral Library (IRTF; \citealt{irtf}) matched to the nearest spectral subtype, and using the planet/star flux ratio to rescale the stellar spectrum to planetary values. For solar- and earlier-type stars, use of an IRTF spectrum had minimal effect compared to stellar spectra approximated as a blackbody. For K-type stars (WASP-43 and TrES-3) the IRTF spectra accounted for molecular absorption and were measurably different from a blackbody, therefore we used IRTF spectra for all spectral types to facilitate comparison.} All planetary spectra were normalized to their continuum flux levels between $1.2-1.3\micron$ for ease of comparison. 

Observations of WASP-33b \citep{wasp33,wasp33confirm} have provided strong evidence for the presence of a thermally inverted atmosphere. When we compare the planetary spectrum of \planet\ to that of WASP-33b binned to similar wavechannels (Fig. \ref{fig:fluxcompare}; top left), we note that the two spectra appear to be very similar in this wavelength range. 
When combined with their additional ground-based, {\it HST}, and {\it Spitzer} data, \cite{wasp33} were able to make a stronger case for a thermally inverted atmosphere than we are able to make with our single measurement of the \planet\ spectrum.

{When compared to the spectrum of WASP-43b \citep{wasp43} and {HD 209458b \citep{line16}}, which have significant water absorption at $1.4\micron$, it becomes clear that the \planet\ spectrum does not display any significant absorption that would be due to ${\rm H_2O}$ (Fig. \ref{fig:fluxcompare}; top center and right, respectively). The decreasing atmospheric profiles of TrES-3b (subsolar ${\rm H_2O}$; \citealt{tres3wasp4}) and HD 189733b (solar ${\rm H_2O}$; \citealt{hd189733b}) adequately match the \planet\ spectrum because of the large scatter and uncertainties in their data (Fig. \ref{fig:fluxcompare}; middle center and right, respectively). However, at the higher S/N of the WASP-4b spectrum \citep{tres3wasp4} the \planet\ spectrum is inconsistent with an ${\rm H_2O}$-depleted decreasing profile (Fig. \ref{fig:fluxcompare}; middle left). }

{The isothermal profile of TrES-3b \citep{tres3wasp4} and the {pseudo}-isothermal profile of CoRoT-2b \citep{corot2b} also closely agree with \planet\ (Fig. \ref{fig:fluxcompare}; bottom left), as does the monotonically decreasing profile of Kepler-13b reported by \cite{beatty16} (Fig. \ref{fig:fluxcompare}; bottom center) and the enhanced C/O-decreasing profile of WASP-12b of \cite{wasp12} (Fig. \ref{fig:fluxcompare}; bottom right), each to within $2\sigma$. These are consistent with our atmospheric matches shown in Fig. \ref{fig:atmos}. That the \planet\ spectrum appears to be similar to those of planets with varying profile shapes indicates that WFC3 observations are not very discriminatory in this wavelength range given the lack of absorption features, and at best indicate a pseudo-isothermal profile.}

The fact that we only see evidence of inversions and (perhaps) TiO absorption in the transmission spectra of the most highly irradiated planets, such as WASP-33b and \planet, is consistent with the hypothesis that cold traps in the interior and on the night side are removing TiO from the atmospheres of more moderate hot Jupiters such as HD 209458b and WASP-43b \citep{spiegel09}. \planet\ will be a key planet for understanding the behavior of TiO in a hot Jupiter atmosphere, and for validating hypothesis about the origin of thermal inversions. Further measurements of \planet\ are necessary towards this effort.

\subsection{Future Work}
\label{sec:future}

Future work on \planet\, should focus on verifying the presence of a thermal inversion in its atmosphere by probing the atmospheric layers in different wavelength regions. Transmission spectra in optical bandpasses would probe atmospheric heights where the isothermal, non-inverted, and thermal inversion models are measurably divergent. 

Optical transmission spectra would also be able to detect potential {absorption} features from TiO or VO, the most likely causes of a thermal inversion in the \planet\ atmosphere. If TiO or VO are present in observable quantities, we could rule out an enhanced C/O atmospheric composition and give more consideration to an inverted atmospheric profile, as an enhanced C/O ratio suppresses formation of TiO and VO \citep{mad11b,mad12}. Detection of IR ${\rm CH_4}$ features would support the existence of an enhanced C/O atmosphere, low TiO and VO levels, and non-inverted atmospheric profile. The {\it Spitzer}  $3.6\micron$ band covers a large $\rm CH_4$ absorption feature, and differencing against a $4.5\micron$ eclipse could measure the relative $\rm CH_4/CO$ levels. 

High-altitude clouds could be identified through optical transmission spectroscopy, which would show a flat transmission spectrum if clouds exist. However, the presence of clouds would likely prevent TiO, VO, or C/O measurements which might allow us to better distinguish between possible atmospheric profiles.

\cite{southworth} observed the transit of \planet\, in Bessell $RI$ and SDSS $griz$ and reported an abnormally steep downward slope in the transmission spectrum from blue to red (see Figure 7 of that paper). The reported slope in the transmission spectrum is too steep to be caused by Rayleigh scattering from haze in the upper atmosphere or by stellar activity \citep{starspot1,starspot2,starspot3}, but could possibly be attributed to TiO {absorption} between ${\sim0.45\micron}$ and $0.8\micron$. \cite{newsouth} also report that the spectral slope measured from their transmission spectroscopy of \planet\, is too strong for Rayleigh scattering and cannot be attributed to effects from the companion star. Additional transit observations in the optical could verify the results of \cite{southworth} and \cite{newsouth}, could lend support to the idea that TiO is present in the \planet\ atmosphere, and help distinguish between the atmospheric models discussed here.

\section{Summary}
\label{sec:summary}

We observed two secondary eclipses of \planet\, from $1.1\micron{\rm\,to\,}1.7\micron$ using the G141 grism on {\it HST}/WFC3 in spatial scan mode. We used Gaussian process regression with MCMC sampling to model both the white-light and spectrally resolved eclipse light curves to extract the planet-to-star flux ratio of the system as a function of wavelength. We corrected this thermal emission spectrum for flux contamination from a nearby star that we probabilistically showed was physically associated with the \wasp\, system. 

We combined the decontaminated thermal emission spectra from each visit of {\it HST} into a visit-averaged spectrum, which was used to retrieve atmospheric models for \planet. After rejecting monotonically decreasing atmosphere models for solar composition and $0.01\times$solar composition, we found that an isothermal or a thermally inverted atmospheric profile could explain our thermal emission spectrum, as could a monotonically decreasing atmosphere with a C/O ratio \textgreater1. We conclude that the \planet\ atmosphere is approximately isothermal across the region probed by our observations, with a brightness temperature of ${\rm T_B=2890}$ K, giving us little power to discern between the fiducial models we tested. 

Additional transit observations in the optical and NIR would test the existence of the steep slope reported by \cite{southworth} and \cite{newsouth}, which would tell us if the atmosphere is truly isothermal or merely pseudo-isothermal as a result of the presence of clouds or haze. A transit spectrum at optical wavelengths would also be able to measure {absorption} from TiO, which, if detected, would favor an inverted atmospheric profile over an enhanced C/O ratio. Alternatively, the detection of IR ${\rm CH_4}$ absorption during secondary eclipse would support an enhanced C/O ratio and disfavor an inverted profile.

\planet, along with other highly irradiated hot Jupiters, will be a key planet for understanding the behavior of TiO in a hot Jupiter atmosphere, and validating hypotheses about the existence and origin of thermal inversions.

\acknowledgements
This work is partially funded by {\it Hubble Space Telescope} grants HST-GO-13660.005 (PI Wright) and HST-GO-13660.001-A (PI Zhao), and partially supported by funding from the Center for Exoplanets and Habitable Worlds. The Center for Exoplanets and Habitable Worlds is supported by the Pennsylvania State University, the Eberly College of Science, and the Pennsylvania Space Grant Consortium. Some of the data presented in this paper were obtained from the Mikulski Archive for Space Telescopes (MAST). STScI is operated by the Association of Universities for Research in Astronomy, Inc., under NASA contract NAS 5-26555. Support for MAST for non-{\it HST} data is provided by the NASA Office of Space Science via grant NNX13AC07G and by other grants and contracts. Some of the data presented herein were obtained at the W.M. Keck Observatory, which is operated as a scientific partnership among the California Institute of Technology, the University of California and the National Aeronautics and Space Administration. The Observatory was made possible by the generous financial support of the W.M. Keck Foundation. The authors wish to recognize and acknowledge the very significant cultural role and reverence that the summit of Mauna Kea has always had within the indigenous Hawaiian community.  We are most fortunate to have the opportunity to conduct observations from this mountain. We gratefully acknowledge the use of SOA/NASA ADS, NASA, and STScI resources. \\ \\

Facilities: \facility{{\it HST} (WFC3)}, \facility{Keck (NIRC2)}.
\\ 





\end{document}